\documentclass[10pt,fleqn]{bmc_article}

\usepackage{cite} 
\usepackage{url}  
\usepackage{ifthen}  
\usepackage{multicol}   
\usepackage[utf8]{inputenc} 
\usepackage{textcomp}
\usepackage{amsmath}
\usepackage{amssymb}

\usepackage{graphics}
\usepackage{float}

\makeatletter

\@fpsep\textheight

\makeatother

\newboolean{publ}
\newenvironment{bmcformat}{\baselineskip20pt\sloppy\setboolean{publ}{false}}{\baselineskip20pt\sloppy}

\begin{document}
\begin{bmcformat}

\title{Light Microscopy: An ongoing contemporary revolution}

\author{Siegfried Weisenburger \email{Siegfried Weisenburger - siegfried.weisenburger@mpl.mpg.de}%
and Vahid Sandoghdar \email{Vahid Sandoghdar - vahid.sandoghdar@mpl.mpg.de}}

\address{%
    Max Planck Institute for the Science of Light and Department of Physics, Friedrich Alexander University of Erlangen-Nuremberg, \\
    91058 Erlangen, Germany
}%
\maketitle

\begin{abstract}
Optical microscopy is one of the oldest scientific instruments that is still used in forefront research. Ernst Abbe's nineteenth century formulation of the resolution limit in microscopy let generations of scientists believe that optical studies of individual molecules and resolving sub-wavelength structures were not feasible. The Nobel Prize in 2014 for super-resolution fluorescence microscopy marks a clear recognition that the old beliefs have to be revisited. In this article, we present a critical overview of various recent developments in optical microscopy. In addition to the popular super-resolution fluorescence methods, we discuss the prospects of various other techniques and imaging contrasts and consider some of the fundamental and practical challenges that lie ahead.
\end{abstract}

\ifthenelse{\boolean{publ}}{\begin{multicols}{2}}{}

\section{Introduction}
The first compound light microscopes constructed in the 16th and 17th centuries enabled scientists to inspect matter and biological specimens at the microscopic level \cite{Toeroek,Wayne}. In 1873, Ernst Abbe formulated a fundamental limit for the resolution of an optical imaging system based on the diffraction theory of light \cite{Abbe1873}. At the same time the fabrication and development of microscopes and lenses were transformed from empirical optimizations to schemes based on quantitative calculations and theoretical considerations. In the 20th century various contrast modalities were developed that allow one to detect very small signals and to measure characteristic properties of a specimen with high specificity. Finally, during the last two decades several revolutionary methods were conceived and experimentally demonstrated, which substantially enhanced the optical resolution down to the nanometer scale (shown in Fig. \ref{fig_opticalresolution}).

The awarding of the 2014 Nobel Prize in Chemistry to Eric Betzig, Stefan Hell and William E. Moerner for their pioneering work in ``super-resolution'' fluorescence microscopy corroborates its promise for many advanced investigations in physics, chemistry, materials science and life sciences.

\begin{figure}[htb]
  \centering
  \includegraphics{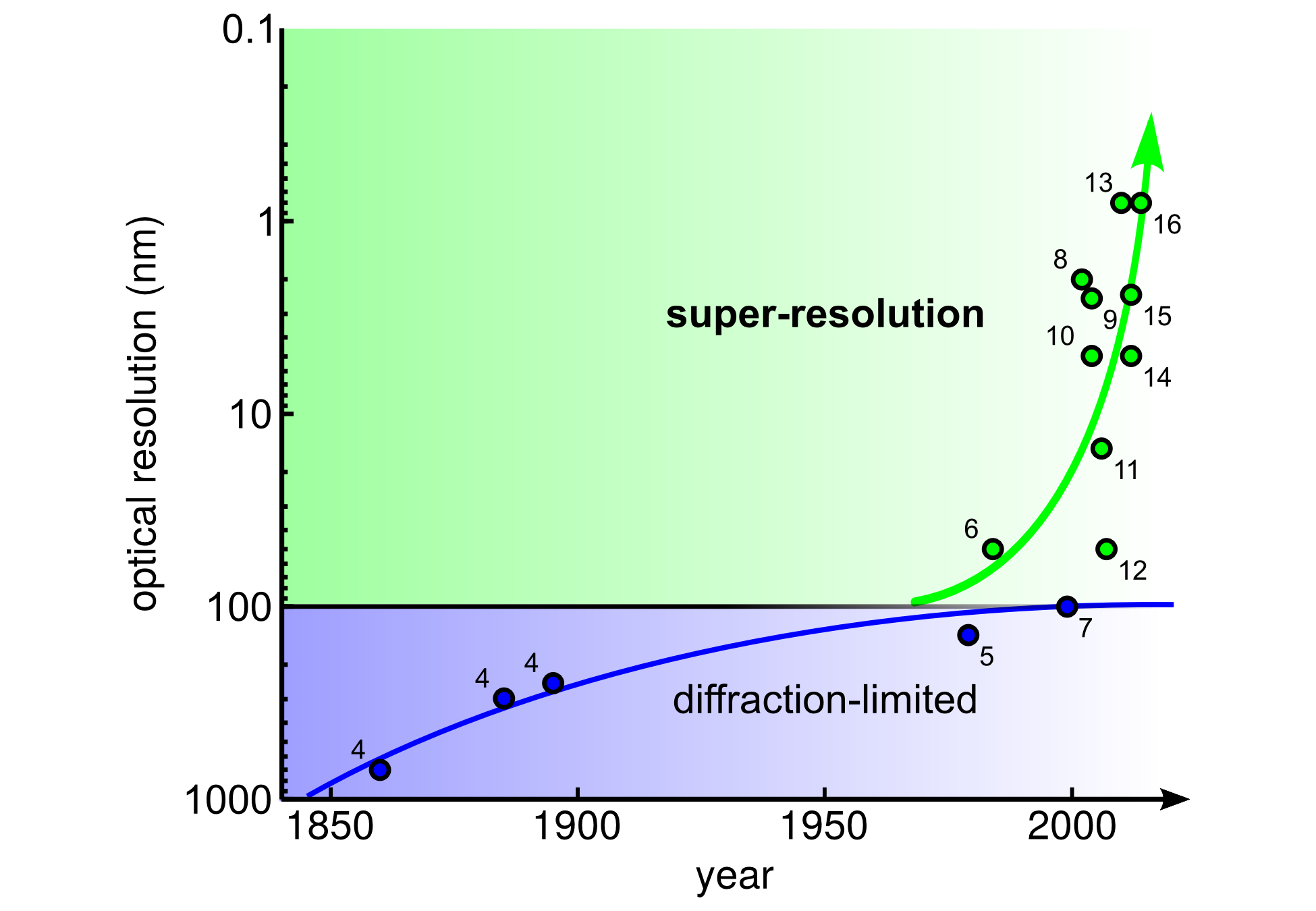}
  \caption{Advancement of the optical resolution over time. Data points are taken from references \cite{OxfordNanotechnology,Cremer1978,Pohl1984,Gustafsson2000,Hettich2002,Qu2004,Gordon2004,Donnert2006,Bretschneider2007,Pertsinidis2010,Vaughan2012,Wildanger2012,SigiCPC2014}.}
  \label{fig_opticalresolution}
\end{figure}

Fluorescence microscopy down to the single molecule level has been reviewed in many recent articles and books \cite{Moerner-Orrit-Wild-Basche,Ishijima2001,Kulzer2004}. Despite the immense success of fluorescence microscopy, this technique has several fundamental shortcomings. As a result, many ongoing efforts aim to conceive alternative modes of microscopy based on other contrast mechanisms. Furthermore, having overcome the dogma of the resolution limit, scientists now focus on other important factors such as phototoxicity and compatibility with live imaging, higher speed, multiscale imaging and correlative microscopy. In this review article, we present a concise account of some of the current trends and challenges.

\section{Ingredients for a good microscope}
\subsection{Contrast}
Every measurement needs an observable, i.e. a signal. In the case of optical microscopy, one correlates a certain optical signal from the sample with the spatial location of the signal source. Scattering is the fundamental origin of the most common signal or contrast mechanism in imaging. Indeed, when one images a piece of stone with our eyes we see the light that is scattered by it although in common language one might speak of reflection. The scattering interaction also leads to a shadow in transmission (see Fig. \ref{fig_transrefl}). In conventional microscopy, one speaks of trans-illumination if one detects the light transmitted through the sample and epi-illumination if one detects the signal in reflection.

\begin{figure}[htb]
  \centering
  \includegraphics{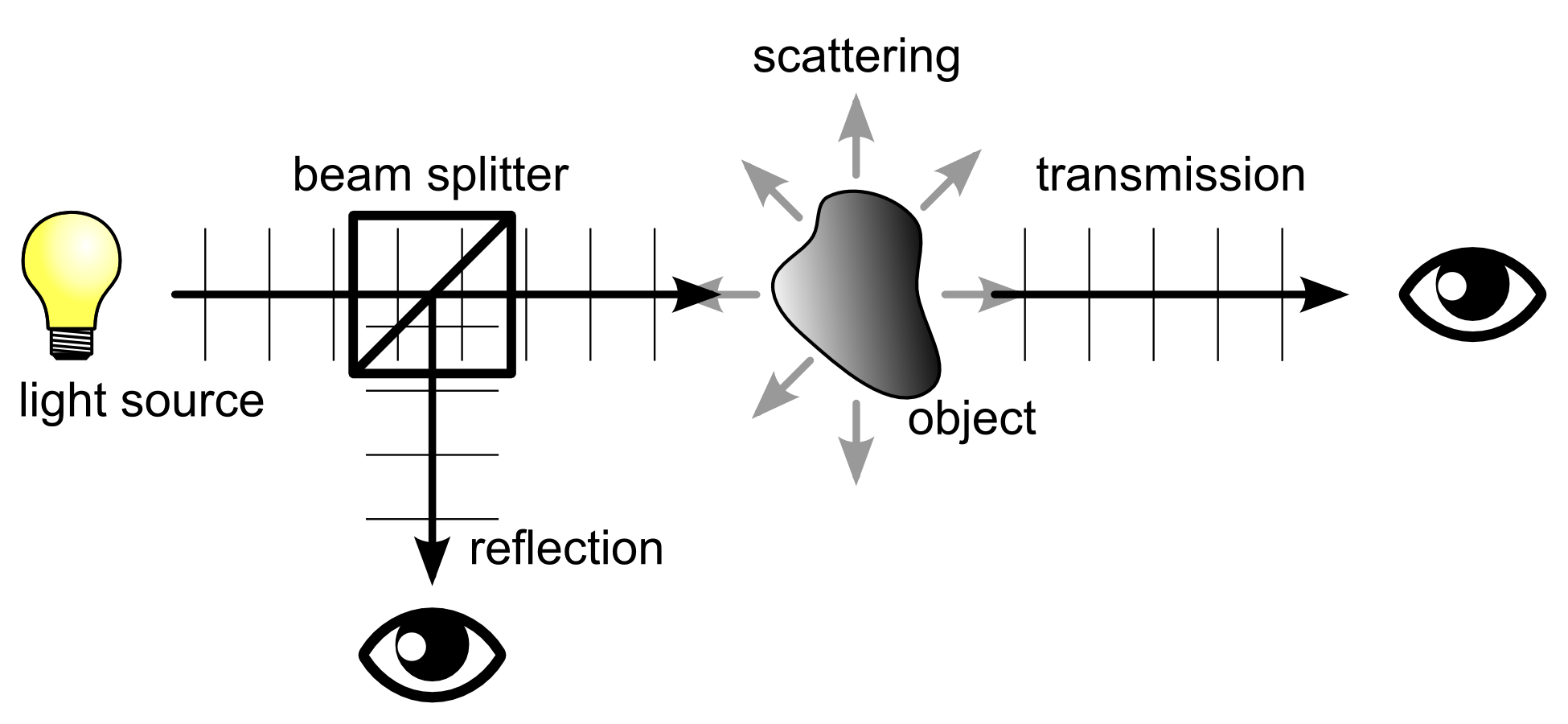}
  \caption{Schematic of a basic arrangement for seeing an object in a microscope. The object is illuminated by a light source, and it is observed either in reflection (epi-illumination) via its scattered light or in transmission (trans-illumination) by looking at its shadow.}
  \label{fig_transrefl}
\end{figure}

Already the pioneers of early microscopy experimented with different types of illumination to generate stronger contrasts. Even today, a good deal of instrumentation in common microscopes focuses on the illumination path. For example, in a particularly interesting scheme one adjusts the illumination angle such that a negligible amount of it is captured by the finite solid angle of the detection optics. In this so-called dark-field microscopy, one emphasizes parts of the sample that scatter light in an isotropic fashion. Such oblique illumination was already exploited in the early days of microscopy \cite{Wayne}. During the past century, various methods were developed to improve the contrast in standard microscopy. For example, polarization microscopy techniques can be used to examine the anisotropy of birefringent materials such as minerals \cite{Talbot1834}.

Some of the most important contrast mechanisms exploit the spectroscopic information of the sample and thus introduce a certain degree of specificity. The prominent example of these is fluorescence microscopy, where fluorophores of different absorption and emission wavelengths are employed to label various parts of a biological species \cite{Pawley}. Over the years, the developments of fluorescence labeling techniques such as immunofluorescence \cite{Coons1950}, engineered organic fluorescent molecules \cite{TsienWaggoner} and fluorescent proteins \cite{Tsien1998} have continuously fueled this area of activity. However, fluorescence labeling has many disadvantages such as photobleaching and most importantly the need for labeling itself.

To address some of the limitations of standard fluorescence imaging, scientists have investigated a range of multiphoton fluorescence methods such as two-photon absorption \cite{Denk1990}. The strong dependence of multiphoton excitation processes on intensity allows one to excite a small volume of the sample selectively only around the focus of the excitation beam. This leads to minimizing the fluorescence background and photobleaching. Aside from its capacity for optical sectioning, this technique makes it possible to perform tissue imaging because the long-wavelength excitation light penetrates deeper into the biological tissue.

The ultimate freedom from fluorescence markers is the use of label-free contrast mechanisms. For example, Raman microscopy generates contrast through the inelastic scattering of light that is selective on the vibrational and rotational modes of the sample molecules \cite{Delhaye1975} and is, thus, very specific to a certain molecule in the specimen. The main difficulty of Raman microscopy lies in its extremely weak cross section. Methods such as coherent anti-Stokes Raman scattering (CARS) \cite{Maker1965} or stimulated Raman scattering \cite{Freudiger2008} improve on the sensitivity to some extent although they remain limited well below the single molecule detection level. Some other interesting label-free contrasts are based on the harmonic generation of the illumination or four-wave mixing processes through the nonlinear response of the sample \cite{Fine1971}. For instance, it has been shown that collagen can be nicely imaged through second harmonic generation (SHG) \cite{Freund1986}.

To conclude this section, we return to the most elementary contrast mechanisms, namely transmission and reflection. The fundamental underlying process in these measurements is interference. Consider again the situation of Fig. \ref{fig_transrefl}, whereby now we reduce the size of the object to a subwavelength scatterer such as a nanoparticle (shown in Fig. \ref{fig_iscat}a). A laser beam with power $P_{\mathrm{inc}}$ illuminates the nanoparticle of diameter $d$ lying on a glass substrate. A fraction of the incoming power $P_{\mathrm{scat}}$ is scattered by the object and another part of it serves as a reference $P_{\mathrm{ref}}$, which can either be the transmitted light or the light that is reflected from the glass interface. The scattered and the reference components of the light reach the detector and interfere if their path difference is much smaller than the coherence length of the illumination beam. In the case of a reflection configuration with field reflectivity $r$, the detected power $P_{\mathrm{det}}$ on the detector is given by
\begin{equation}
  \label{interference}
  P_{\mathrm{det}} = P_{\mathrm{inc}} \left( r^2 + s^2 + 2rs\cos{\varphi} \right) = P_{\mathrm{ref}} + P_{\mathrm{scat}} + P_{\mathrm{inter}} \quad ,
\end{equation}
where $s^2$ signifies the scattering cross-section of the particle and $\varphi$ denotes a phase term that includes the Gouy phase \cite{Gouy1890,SalehTeich}. Interferometric scattering contrast was explicitly formulated in the context of detection of nanoscopic particles in our laboratory \cite{Lindfors2004} and has been coined iSCAT \cite{Kukura2009b}. The central process is, however, quite general and related to the measurement of extinction by an object on a beam of light. According to the Optical Theorem, the extinction signal can stem from scattering loss or absorption \cite{Bohren-Huffman}. We note that although the interaction of light with objects that are larger than a wavelength is expressed in terms of reflection instead of scattering, the signal on the detector can still be written in the form of Eq. (\ref{interference}).

Having understood the underlying interferometric nature of an extinction measurement, it is now clear that one can also perform the measurement by manipulating the reference beam in Eq. (\ref{interference}). An early example of such a variation was demonstrated by Zernike in the context of phase contrast microscopy \cite{Zernike1955}. Here a part of the illumination is phase shifted and then mixed with the light that is transmitted through the sample. Another related technique is known as differential interference contrast microscopy (DIC) put forth by Nomarski \cite{Nomarski1955}. A very similar method to DIC that is somewhat simpler to implement and therefore very common in commercial microscopes is Hoffman modulation contrast \cite{Hoffman1975}. Although simple transmission, reflection, phase contrast and DIC share the same fundamental contrast as that of iSCAT, only the latter has tried to address the issue of sensitivity and its extension to very small nanoparticles and single molecules. This brings us to the general question of sensitivity.

\subsection{Sensitivity}
For any given contrast mechanism, one can ask ``how small of an object can one detect?''. In other words, what is the sensitivity. To have high sensitivity, one needs a sufficiently large signal from the object of desire, and the trouble is that usually the signal diminishes quickly as the size of the object is reduced. So, one needs to collect the signal efficiently and employ very sensitive detectors. The detector can either be a point detector as it is most often used in scanning confocal techniques or a camera in the case of wide-field imaging. The performance of a light detector can be generally described by its quantum efficiency, the available dynamic range and its time resolution \cite{Jung2013}.

For the longest time in the history of microscopy the human eye was the only available detector. With its detection threshold of only a few photons it was a better detector than photographic plates and films even long after these became available \cite{Rieke1998}. Photon counting detectors had emerged by the 1970s, starting with photo-multiplier tubes (PMT) \cite{Cova1973} and later followed by semiconductor devices that are able to detect single photons. These single-photon avalanche diodes (SPAD) can nowadays achieve quantum efficiencies above 50 \% with timing resolutions on the order of tens of picoseconds. In the 1990s, cameras like the charge-coupled device (CCD) and fast active pixel sensors using complementary metal-oxide-semiconductor (CMOS) technology reached the single-photon sensitivity with high quantum efficiencies. Today, the best CCD cameras using electron multiplication can achieve quantum efficiencies better than 95 \% in the visible part of the spectrum with a readout noise of a fraction of a photo-electron.

\begin{figure}[htb]
  \centering
  \includegraphics{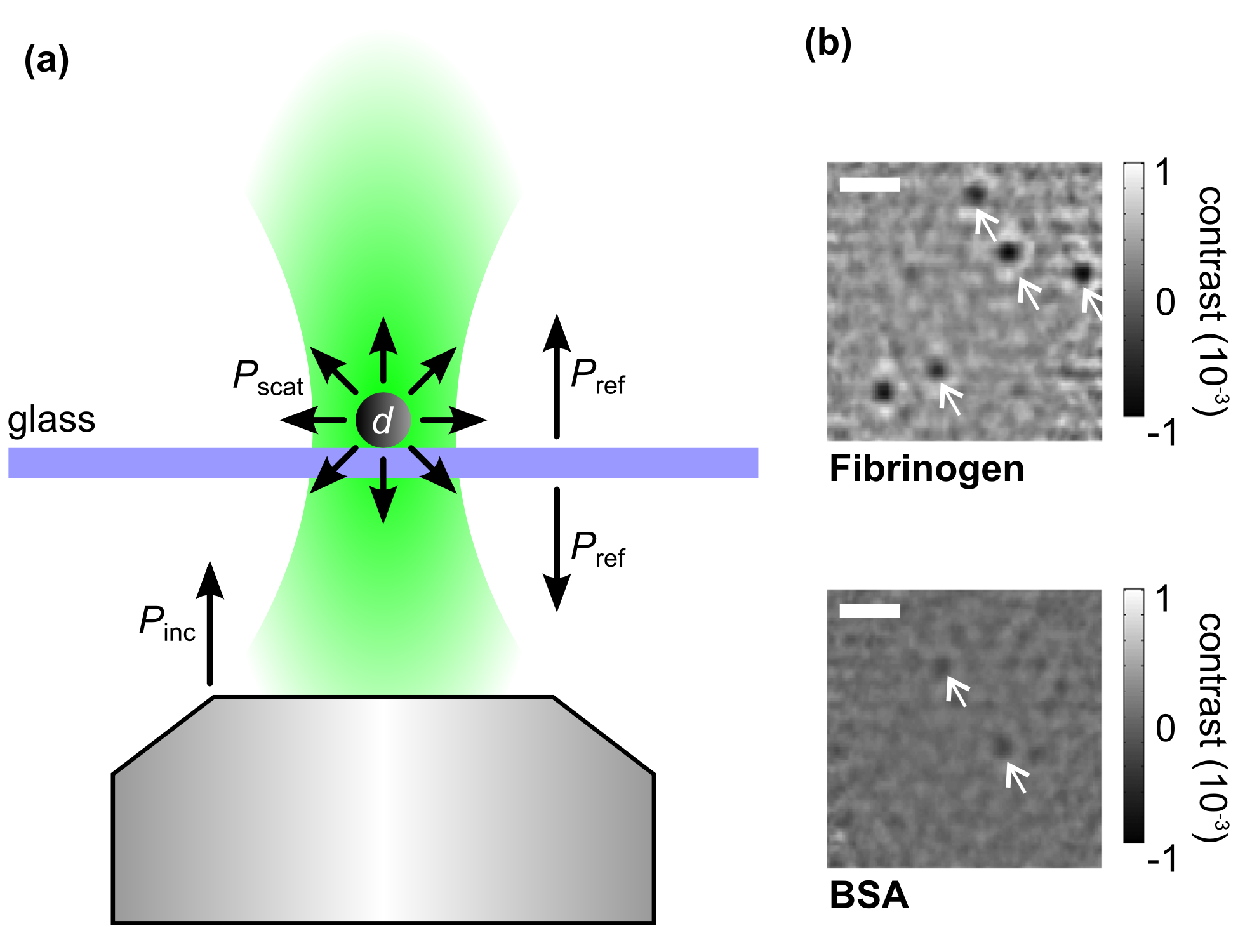}
  \caption{Interferometric scattering detection (iSCAT). {\bfseries (a)} Principle of iSCAT in reflection configuration. A nanoparticle of diameter $d$ is illuminated by a light beam with power $P_{\mathrm{inc}}$. Part of the light ($P_{\mathrm{ref}}$) will serve as a reference beam and part of the light is scattered by the particle ($P_{\mathrm{scat}}$). In the imaging plane the two beams can interfere leading to an enhancement of the scattering signal. {\bfseries (b)} Examples of differential iSCAT images are shown for two protein types. Individual molecules are marked with arrows. Scale bars: $1 \mu m$. (b: Modified with permission from \cite{Piliarik2014}.)}
  \label{fig_iscat}
\end{figure}

An additional important issue regarding detection sensitivity concerns the background signal, which can swamp the signal. If the detector dynamic range (the ratio of the highest and the smallest possible amount of light that can be measured) is large enough, one can subtract the background. However, temporal fluctuations on the background usually limit this procedure in practice. In general, the image sensitivity can be quantified in terms of the signal-to-noise ratio (SNR),
\begin{equation}
  \label{eqn_sensitivity}
  \mathrm{SNR} = \frac{\mu(\mathrm{signal}) - \mu(\mathrm{background})}{\sqrt{\sigma^2(\mathrm{signal}) + \sigma^2(\mathrm{background})}} \quad ,
\end{equation}
where $\mu$ denotes the mean and $\sigma$ the standard deviation.

The sensitivity of fluorescence microscopy was taken to the single-molecule limit towards the end of the early 1990s \cite{Orrit1990,Betzig1993,Nie1994}. The factors leading to the success of this field were 1) access to detectors capable of single-photon counting, 2) the ability to suppress the background by using spectral filters, 3) preparation of clean samples. Although single-molecule fluorescence microscopy has enabled a series of spectacular studies in biophysics, fluorescence blinking and bleaching as well as low fluorescence quantum yield of most fluorophores pose severe limits on the universal applicability of this technique.

The low cross sections of Raman and multiphoton microscopy methods also hinder the sensitivity of these methods. Only in isolated cases, where local field enhancement has been employed, have been reports of single-molecule sensitivity \cite{Nie1997,Stiles2008}. For a long time, single-molecule sensitivity in extinction was also believed not to be within reach because the extinction contrast of a single molecule is of the order of $10^{-6}$, making it very challenging to decipher the signal on top of laser intensity fluctuations. Nevertheless, various small objects have been detected and imaged using iSCAT, including very small metallic nanoparticles \cite{Lindfors2004,Jacobsen2006}, single unlabeled virus particles \cite{Kukura2009b}, quantum dots even after photobleaching \cite{Kukura2009}, single molecules \cite{Kukura2010,Celebrano2011} and even single unlabeled proteins down to a size of 60 kDa \cite{Piliarik2014} (shown in Fig. \ref{fig_iscat}b). Here, it is important to note that the power that is scattered by a Rayleigh particle ($P_{\mathrm{scat}}$) follows a $d^6$ law where $d$ denotes the particle diameter. The interference term $P_{\mathrm{interference}}$ (cf. Eq. \ref{interference}) contains the scattering field or in other words the polarizability of the particle, which is proportional to $d^3$. Therefore, this term will dominate the detected power for very small objects. The high sensitivity of iSCAT means, however, that any slight variation in the index of refraction or topography can lead to a sizable contrast. Hence, it is important to account for fluctuations of the index of refraction, length or absorption in the sample.

\subsection{Resolution}
One of the most immediate functions that the layperson associates with a microscope is its ability to reveal the small features of a sample. The principle of operation of a microscope is typically described using ray optics. However, when dimensions are to be investigated of the order of the wavelength of visible light, i.e. 400 - 800 nm, we must consider the wave properties of light such as interference and diffraction \cite{Abbe1873}. Therefore one cannot achieve an arbitrarily high resolution simply by increasing the magnification of the lens arrangement.

Ernst Abbe pioneered a quantitative analysis of the resolution limit of an optical microscope \cite{Abbe1873}. He considered imaging a diffraction grating under illumination by coherent light \cite{Cremer2013}. Abbe argued that one would recognize the grating if one could detect at least the first diffraction order (see Fig. \ref{fig_psf}a). In the case of an immersion microscope objective with circular aperture and direct on-axis illumination, the Abbe diffraction limit of resolution $d$ reads
\begin{equation}
  \label{eqn_abbe}
  d = \frac{\lambda}{\mathrm{NA}} \quad ,
\end{equation}
where $\lambda$ is the wavelength of light and $\mathrm{NA} = n \sin{\alpha}$ was introduced as the numerical aperture (illustrated in Fig. \ref{fig_psf}b). Here $n$ is the refractive index of the medium, in which the microscope objective is placed, and $\alpha$ denotes the half-angle of the light cone that can enter the microscope objective. Air has a refractive index of about $n = 1$ limiting the $\mathrm{NA}$ of dry microscope objectives to less than unity. By filling the space between cover glass and an immersion microscope objective with a high index material, the numerical aperture can be increased.

Already in the original publication, Ernst Abbe discussed how this diffraction-limited resolution can be improved if the illumination comes at an angle with respect to the optical axis, making it possible to collect higher diffraction orders (cf. Fig. \ref{fig_psf}a). In this case, the diffraction limit is determined by the sum of the numerical apertures of the illumination lens and the collection lens. If the angles of incidence and collection are identical, a factor of $2$ is obtained, leading to the famous Abbe formula
\begin{equation}
  \label{eqn_incoherent}
  d = \frac{\lambda}{2 \mathrm{NA}} \quad.
\end{equation}
Considering that the optical response of an arbitrary object can be Fourier decomposed, Abbe's formula can be used as a general criterion for resolving its spatial features.

At about the same time, Hermann von Helmholtz developed a more elaborate mathematical treatment that he published one year after Abbe \cite{Helmholtz1874}. Helmholtz also discussed incoherent illumination and showed that Eq. (\ref{eqn_incoherent}) holds in that case too. In a nutshell, for an incoherent illumination, the light is scattered from the individual nanoscopic parts of the structure rather than diffracted in a well-defined order. Thus one can consider the illumination to be from every direction \cite{GoodmanFourier}.

Lord Rayleigh was the first to consider self-luminous objects, which also emit incoherently \cite{Rayleigh1896}. Additionally, he discussed different types of objects, different aperture shapes and the similarities in the diffraction limit for microscopes and telescopes. Although Abbe's resolution criterion is more rigorous, a more commonly known formulation of the resolution for spectrometers and imaging instruments is the Rayleigh criterion,
\begin{equation}
  \label{eqn_rayleigh}
  d = 1.22\frac{\lambda}{2 \mathrm{NA}} \quad.
\end{equation}

It states that two close-lying points are considered resolved if the first intensity maximum of one diffraction pattern described by an Airy disc \cite{Airy1835} coincides with the first intensity minimum of the other diffraction pattern \cite{Rayleigh1879}. Equation (\ref{eqn_rayleigh}) is somewhat arbitrary but it comes very close to Abbe's criterion. Rayleigh based his definition upon the physiological property of the human eye, which can only distinguish two points of a certain intensity difference.

In practice, the full-width at half-maximum (FWHM) of the point-spread function (PSF) also provides a useful criterion for the resolution of a microscope because two overlapping PSFs that are much closer than their widths can no longer be resolved. For an immersion oil microscope objective with $\mathrm{NA} = 1.49$ operating at a wavelength of 500 nm, the PSF is about 220 nm (see Fig. \ref{fig_psf}b). The axial width of the PSF is about 2-3 times larger, in this case amounting to about 400 nm.

\begin{figure}[htb]
  \centering
  \includegraphics{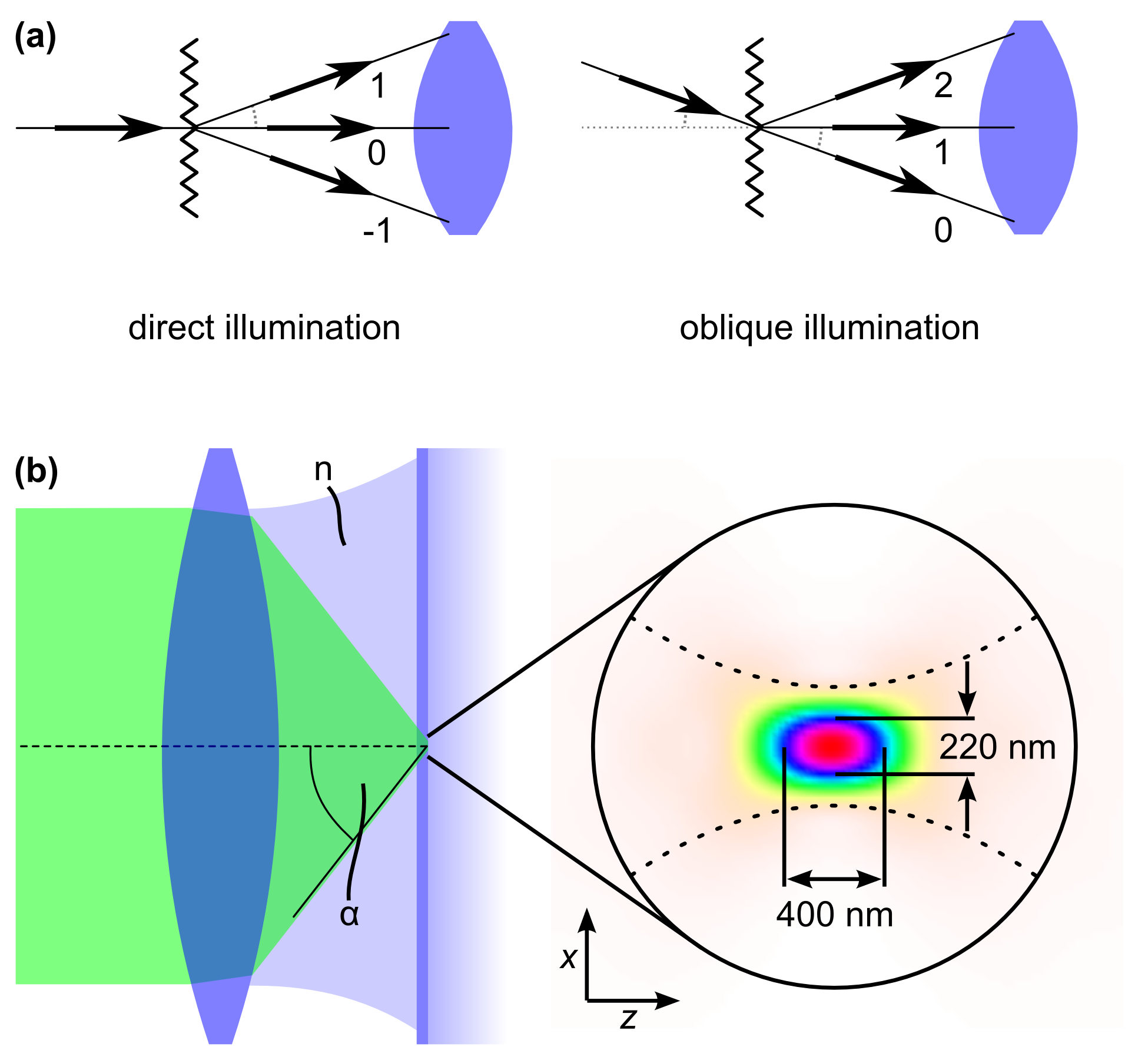}
  \caption{Resolution and point-spread function of an optical microscope. {\bfseries (a)} Comparison of direct and oblique illumination for Abbe's considerations. In the case of oblique illumination higher diffraction orders can be collected by a lens with the same NA. {\bfseries (b)} Diffraction-limited point-spread function of an oil immersion microscope objective. Calculated for wavelength $\lambda = 500$ nm and a numerical aperture NA = 1.49. The lateral and axial FWHMs of the PSF amount to about 220 nm and 400 nm, respectively.}
  \label{fig_psf}
\end{figure}

It is worth mentioning that the PSF can also take different forms depending on the employed optical beam. For example, it has been shown that the PSF of a focused doughnut beam with radial polarization can be made somewhat smaller than that of a conventional TEM$_00$ mode \cite{Dorn2003}. The origin of this effect is the vectorial character of optical beams.

\section{Improving the resolution in fluorescence imaging}
Fluorescence is one of the most important contrast mechanisms because it offers the possibility of specific labeling. However, since the spontaneous emission of a fluorophore does not preserve the coherence of the illumination, the signal is incoherent. As we shall see shortly, one can engineer the illumination to increase the resolution by a factor of $2$.

\subsection{Confocal scanning microscopy}
One of the possibilities for improving the resolution is confocal microscopy. Although the principle was patented by Marvin Minsky in 1957, it took 20 years until the invention of suitable lasers and the progress in computer-controlled data acquisition opened the door for its widespread use \cite{Cremer1978}. In contrast to conventional wide-field illumination, where the full field of view is illuminated and imaged onto a camera, scanning confocal microscopy uses spatial filtering to ensure that only a diffraction-limited focal volume is illuminated and that only light from this focal volume can reach the detector. An image is then produced by raster scanning this confocal volume across the sample. Since out-of-focus light is effectively suppressed, the method allows for higher contrast and offers the ability to perform optical sectioning to acquire 3D images.

The lateral size of the PSF can be improved by a factor of $\sqrt{2}$ in confocal microscopy \cite{Sheppard1977}. However, in reality this factor depends on the coherence properties of the imaging light and the finite size of the detection pinhole \cite{Sheppard1977}. The latter is usually set to a value about the size of the point spread function so as to not lose any signal. As a result, only a marginal improvement of the resolution can be achieved in confocal microscopy.

A particularly interesting mode of confocal microscopy is image scanning microscopy, where an image is recorded on a camera at each scan point \cite{Sheppard1988,Mueller2010}. It has been shown that one can computationally reconstruct an image with a resolution that is improved by up to a factor of $2$ from the resulting image stack.

\subsection{Structured illumination microscopy (SIM)}
The scanning feature of confocal microscopy limits its speed and, thus, wide-field approaches are generally favored. An attractive and powerful strategy to improve the lateral resolution of wide-field fluorescence microscopy by up to a factor of $2$ is offered by structured-illumination microscopy (SIM) \cite{Gustafsson2000}. Here, the sample is illuminated using a patterned light field, typically sinusoidal stripes produced by interference of two beams that are split by a diffraction grating. The resulting image is a product of the illumination pattern and the fluorescence image of the sample. Assuming that the dye concentration follows a certain pattern that can be Fourier decomposed, one obtains a Moir\`e pattern for each component, resulting from the product of the array of illumination lines and the fluorescence signal (Fig. \ref{fig_sim}a). The key concept in SIM is that the periodicity of a Moir\`e pattern is lower than the individual arrays.

Let us consider a sample with a periodic array of dyes at distance $a_s$ illuminated by an array of lines spaced by $a_i$. If we take the angle of the two line arrays to be zero, the period $a_M$ of the Moir\`e pattern is given by $a_M = (a_s \cdot a_i)/\lvert a_s-a_i \rvert$. Now, consider an objective lens that accepts spatial frequencies $k_o$. The highest periodicity Moir\`e pattern that is detected by this objective is $a_M = 1/k_o$, so that the decisive criterion becomes
\begin{equation}
  \label{eqn_oursim}
  \frac{1}{a_M} = \left| \frac{1}{a_i} - \frac{1}{a_s} \right| \le k_0 \quad.
\end{equation}
Furthermore, considering that the highest periodicity illumination that is compatible with the objective is $a_i = 1/k_o$. Putting all this information together, one finds that it is possible to detect Fourier components of the sample at $2k_o = 1/a_s$. In other words, one can resolve sample features at a spatial frequency up to twice larger than usual.

\begin{figure}[htb]
  \centering
  \includegraphics{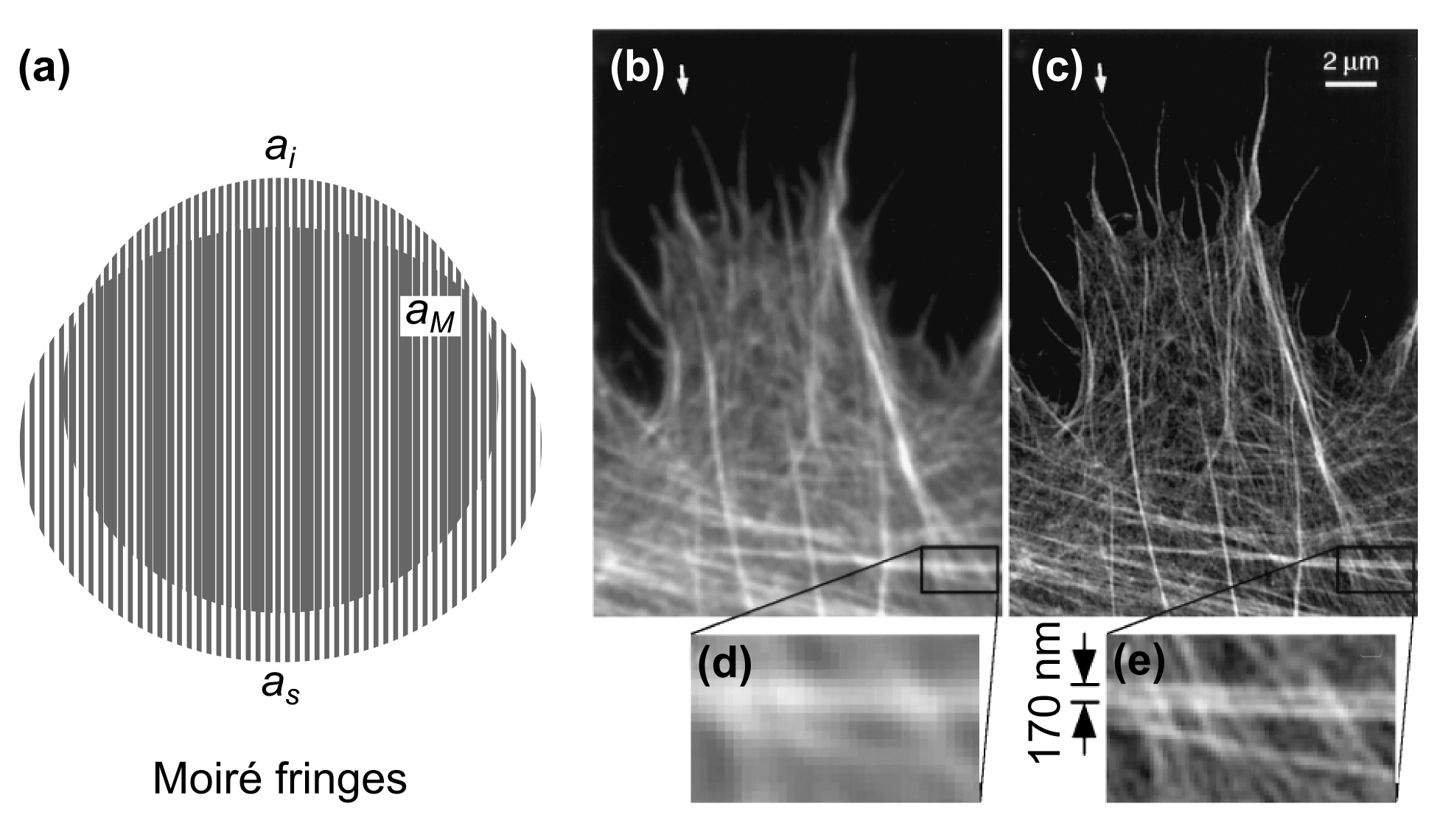}
  \caption{Structured illumination microscopy (SIM). {\bfseries (a)} Two fine structures (the known illumination pattern and the unknown sample structure) produce a Moir\`e interference pattern that is imaged by the microscope. The method allows for an improvement of factor $2$ in resolution; see text for details. {\bfseries (b,c)} Actin cytoskeleton at the edge of a HeLa cell imaged by conventional microscopy and SIM, respectively. {\bfseries (d,e)} Insets show that the widths (FWHM) of the finest protruding fibers (small arrows) are lowered to 110 - 120 nm in {\bfseries (c)}, compared to 280 - 300 nm in {\bfseries (b)}. (b-e: Reproduced with permission from \cite{Gustafsson2000}.)}
  \label{fig_sim}
\end{figure}

By recording a series of images for different orientations and phases of the stripe pattern, one can reconstruct the full image with an improved resolution. Interestingly, the resulting Moir\`e pattern has also a three-dimensional structure, which allows the reconstruction program to obtain an enhanced resolution in the axial direction. Figure \ref{fig_sim}c shows an example of a high resolution image obtained by SIM and the comparison to conventional microscopy (Fig. \ref{fig_sim}b).

There are also approaches that combine structured illumination with other techniques using interference and two opposing objectives to gain high axial resolution (I$^n$M, $n = 2,3,5$) \cite{Gustafsson1999,Shao2008}. Recently, SIM has also been combined with light-sheet microscopy (coined lattice light-sheet microscopy) \cite{Chen2014}. In an intriguing demonstration, 3D in-vivo imaging of relatively fast dynamical processes is shown using a bound optical lattice as the light sheet. It is worth mentioning that SIM strategies cannot yield an extra factor of two improvement for coherent imaging modalities \cite{Wicker2014}.

\section{Near-field microscopy}
\subsection{Scanning near-field optical microscopy}

In 1928, Edward Synge proposed to use a thin opaque metal sheet with sub-wavelength holes to illuminate a sample placed at sub-wavelength separation from it \cite{Synge1928}. By scanning the specimen through this point illumination, an image could be recorded with an optical resolution better than the diffraction limit. Technical limitations in the fabrication of nanoscopic apertures, their nano-positioning and sensitive light detection made the experimental realization of a scanning near-field optical microscope (SNOM) only possible in the early 1980s \cite{Pohl1984}.

\begin{figure}[htb]
  \centering
  \includegraphics{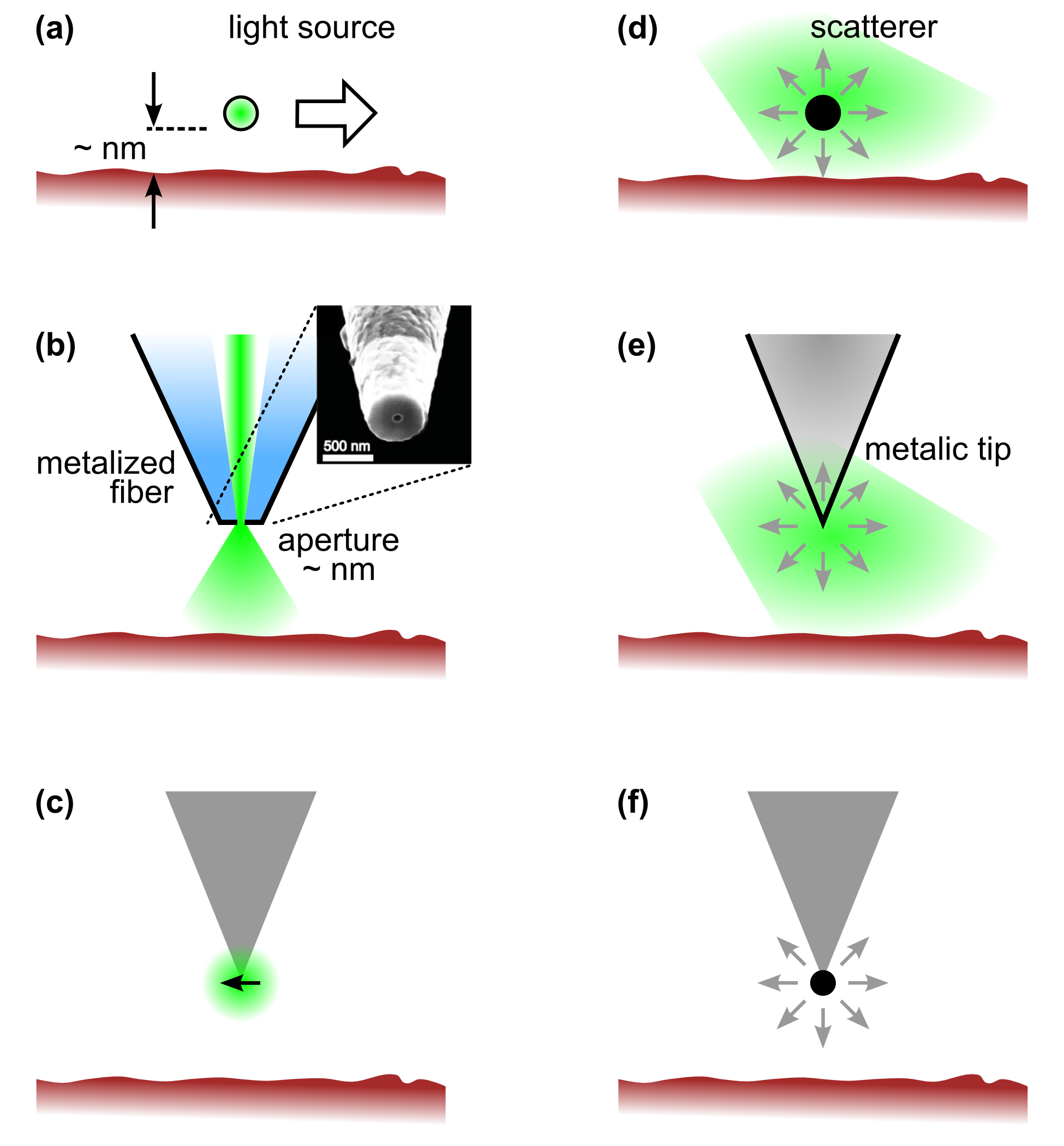}
  \caption{Scanning near-field optical microscopy. {\bfseries (a)} A nanoscopic light source is brought very close to the specimen, typically at a height of a few nanometers, and it is raster scanned across the surface of the sample. {\bfseries (b)} Aperture SNOM using a metalized optical fiber tip. Inset: Electron microscope image of an aperture SNOM probe. {\bfseries (c)} Schematics of SNOM with a single molecule or color center as a light source mounted at the end of a tip. {\bfseries (d)} The light source is replaced by a nanoscopic scatterer, which is illuminated in a conventional far-field fashion. Both in (a) and (b), the detection is done in the far field. {\bfseries (e)} An apertureless SNOM using a solid tip. {\bfseries (f)} Schematics of an ideal arrangement for apertureless SNOM using a single well-defined nanoscopic scatterer.}
  \label{fig_nearfield}
\end{figure}
Near-field microscopy gets around the diffraction limit in a complete and general fashion. The essential point is that the limitations imposed by diffraction do not apply to the distances very close to the source, where non-propagating evanescent fields dominate. These fields contain the high spatial frequency information of the source and sample, but their intensity decays exponentially with a characteristic length of the order of the wavelength of light. There are two conceptual ways of performing SNOM. In the first case, one follows Synge's proposal and illuminates the sample through a subwavelength light source (see Fig. \ref{fig_nearfield}a). Conventionally, the light source is realized by sending light through a tapered fiber that is metalized and has a subwavelength aperture at its end (see Fig. \ref{fig_nearfield}b). This mode of operation is called aperture SNOM. Here the size and shape of the aperture dictate the range of spatial frequencies that can be coupled to and scattered by the sample. The detection can be performed in the far field or back through the aperture although the latter leads to a very unfavorable signal-to-background ratio. The main limitations of this arrangement are 1) small transmission of $10^{-3}-10^{-5}$ through the aperture, and 2) the fundamental limit of aperture given by the skin depth of metals \cite{Novotny1995}.

In practice, the nanoscopic light source, aperture or scatterer is placed at the end of a sharp tip and various distance regulation mechanisms such as the shear force are used to raster scan it at nanometer separation from the sample. The most common realization of the aperture SNOM employs a metal-coated tapered optical fiber with a small aperture of the order of a few nanometers at its apex (see Fig. \ref{fig_nearfield}b). However, as mentioned earlier, reaching a resolution below 50 nm proves to be exceedingly hard due to the low tip throughput and the fact that the effective size of the aperture cannot be reduced below about 20 nm \cite{Novotny1995}. In principle, this problem can be circumvented by replacing the aperture with a nanoscopic source of light such as a single molecule or a single color center \cite{Michaelis2000, Kuehn2001} (Fig. \ref{fig_nearfield}c). However, the difficulties of placing the emitter at the very end of the tip and its photostability at room temperature have hampered the wide-spread adoption of this method.

In the second SNOM approach, one uses far-field illumination and scans a nanoscopic scatterer or antenna very close to the sample (cf. Fig. \ref{fig_nearfield}d). The near field of this nanostructure again contains high spatial frequencies, which can be scattered by the sample. In general, this so-called apertureless SNOM mode \cite{Zenhausern1994} is much more challenging. The main difficulty is that the far-field illumination creates a large background from an area (minimum of a diffraction-limited area) that is much larger than the near-field domain of the antenna. Furthermore, this background light easily results in interferometric artifacts when the height of the antenna or sample are changed \cite{Hecht1997,Sandoghdar1997}.

The most common platform for performing apertureless SNOM has been the solid tip, which might be metalized \cite{Zenhausern1995, Kawata1994, Gleyzes1995, Knoll1999} (Fig. \ref{fig_nearfield}e). Although such experiments in the infrared domain have become well established \cite{Hillenbrand2004}, reproducible fabrication and efficient characterization of the suitable tips have made apertureless SNOM in the visible regime an uphill battle. As a result, many scientists turned to using more well-defined metallic nanoparticles placed at the end of a dielectric tip\cite{Kalkbrenner2001} (Fig. \ref{fig_nearfield}f). This approach has been particularly successful in the context of antenna-based SNOM and in producing quantitative data on the near-field enhancement of fluorescence from single molecules \cite{Kuehn2006,Anger2006,Lee2007,Hoeppener2008,Eghlidi2009,Wientjes2014}.

\subsection{Wide-field near-field optical microscopy}
An alternative way of imaging in the near field has emerged in the context of metamaterials. These artificial materials are structures with sub-wavelength-sized unit cells and can be engineered to exhibit intriguing properties such as a negative index of refraction \cite{Veselago1968,Valentine2008}. In the year 2000, a so-called perfect lens was proposed by John Pendry using a slab of negative index material \cite{Pendry2000}. As illustrated in Fig. \ref{fig_nearfield2}a, the idea with such a lens is that an object that is embedded in a material with refractive index $n = 1$ can be perfectly imaged by a slab of a material with refractive index $n = -1$, assuming that it is perfectly impedance matched and completely lossless. The exponential decay of the evanescent wave intensity inside the $n = 1$ material is completely reversed by an exponential amplification inside the material with $n = -1$ refractive index. Thus both propagating and evanescent fields are imaged by this lens yielding a perfect image.
\begin{figure}[htb]
  \centering
  \includegraphics{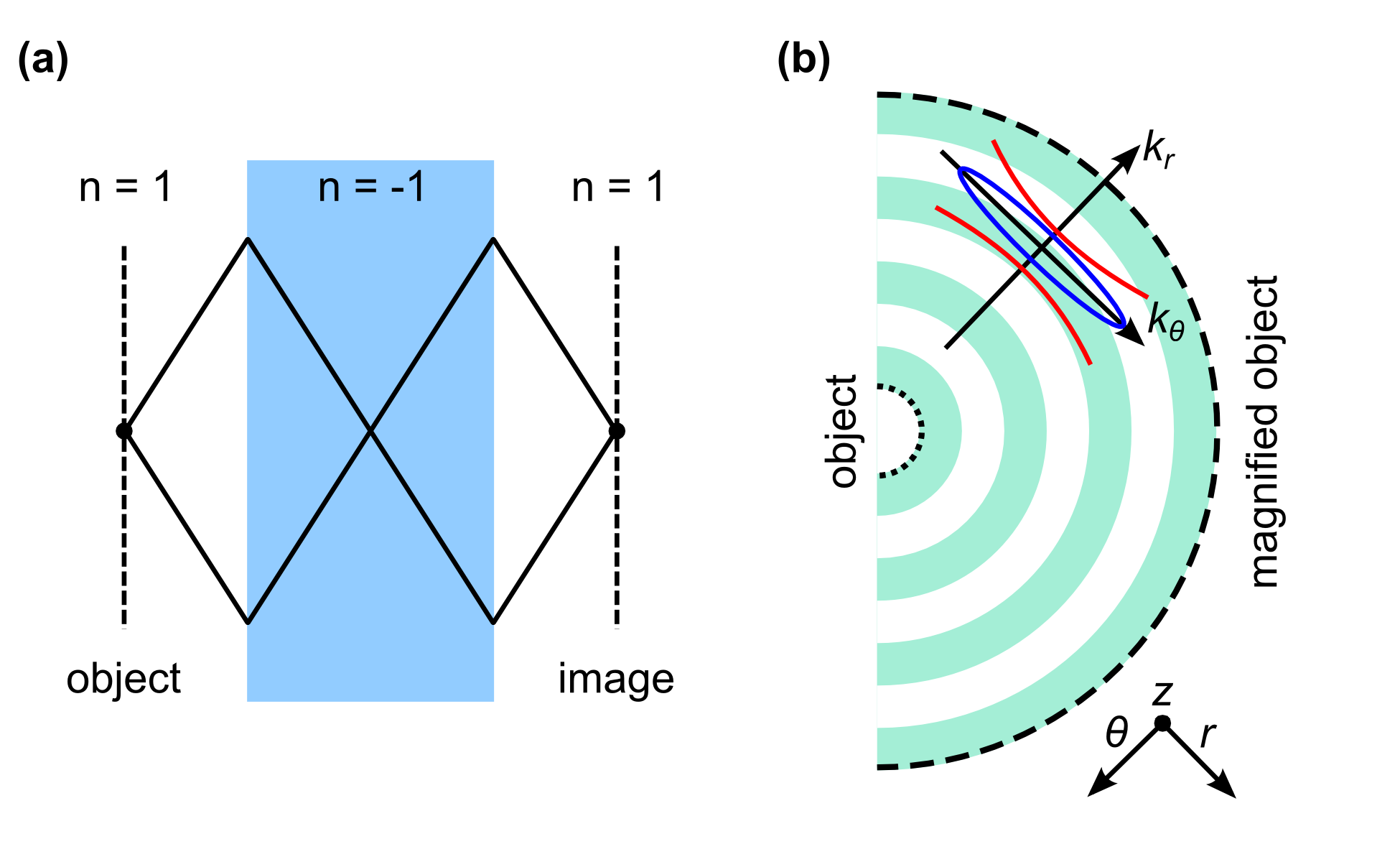}
  \caption{Wide-field near-field optical microscopy. {\bfseries (a)} Illustration of a perfect lens using metamaterials. {\bfseries (b)} Schematic of a cylindrical hyper lens. The dispersion properties of a multilayer metamaterial with alternating dielectric and metallic layers are engineered to have hyperbolic or eccentric elliptic dispersion. For a sub-wavelength object placed on the inside layer, wave propagation along the radial direction gradually compresses the tangential wavevectors resulting in a magnified image at the outer boundary that can be imaged using conventional optics.}
  \label{fig_nearfield2}
\end{figure}

Such a perfect lens has not yet been experimentally realized because of the extremely delicate constraints on the properties of the negative index material. There are, however, experimental demonstrations of a superlens in the optical regime where sub-diffraction imaging was shown using metamaterial structures \cite{Fang2005}. There have also been efforts for the realization of a hyperlens (see Fig. \ref{fig_nearfield2}b) to project the near field into the far field using a cylindrical \cite{Liu2007} or a spherical hyperlens \cite{Rho2010}. Fabrication issues, material properties and the requirement that the object of interest must be placed in the near field of the hyperlens, make the practical usage of these interesting imaging techniques very limited although they might possibly find use in nanofabrication and optical data storage.

\section{Far-field super-resolution fluorescence microscopy}

\subsection{STED and STEDish techniques}
Fluorescence microscopy witnessed several important developments in the 1980s. This included commercialization of scanning confocal microscopy and the invention of two-photon absorption microscopy. A young doctoral student in Heidelberg, Stefan Hell, took upon himself not to accept the diffraction limit in its usual formulation. Shortly after finishing his doctoral work, Hell proposed to exploit stimulated emission to deplete (STED) the fluorescence of molecules in the outer part of the illumination and thereby reduce the size of the effective fluorescence spot \cite{Hell1994}. The first experimental realization of this idea was reported just a few years later \cite{Klar1999}. Hell was awarded the Nobel Prize in Chemistry in 2014 for his achievements in this area.

\subsubsection{Stimulated emission depletion}
Upon excitation, a fluorescent molecule is usually brought from its singlet ground state S$_0$ to a higher vibrational state of the singlet electronic state S$_1$ (see Fig. \ref{fig_sted}a), which then relaxes on a picosecond time scale to the lowest vibrational level. If the quantum efficiency of the molecule is high, a photon is emitted within a few nanoseconds. Hell had the idea to suppress this fluorescence in the outer part of the excitation beam by stimulating the emission much faster than the nanosecond spontaneous emission in that region. To do this, he used a doughnut-shaped laser beam profile (see Fig. \ref{fig_sted}b) that is overlapped on the excitation focal spot. The stimulated emission takes place at a wavelength that is red-shifted with respect to the main part of the fluorescence line, allowing one to spectrally separate the stimulated emission from various fluorescence components.

\begin{figure}[H]
  \centering
  \includegraphics{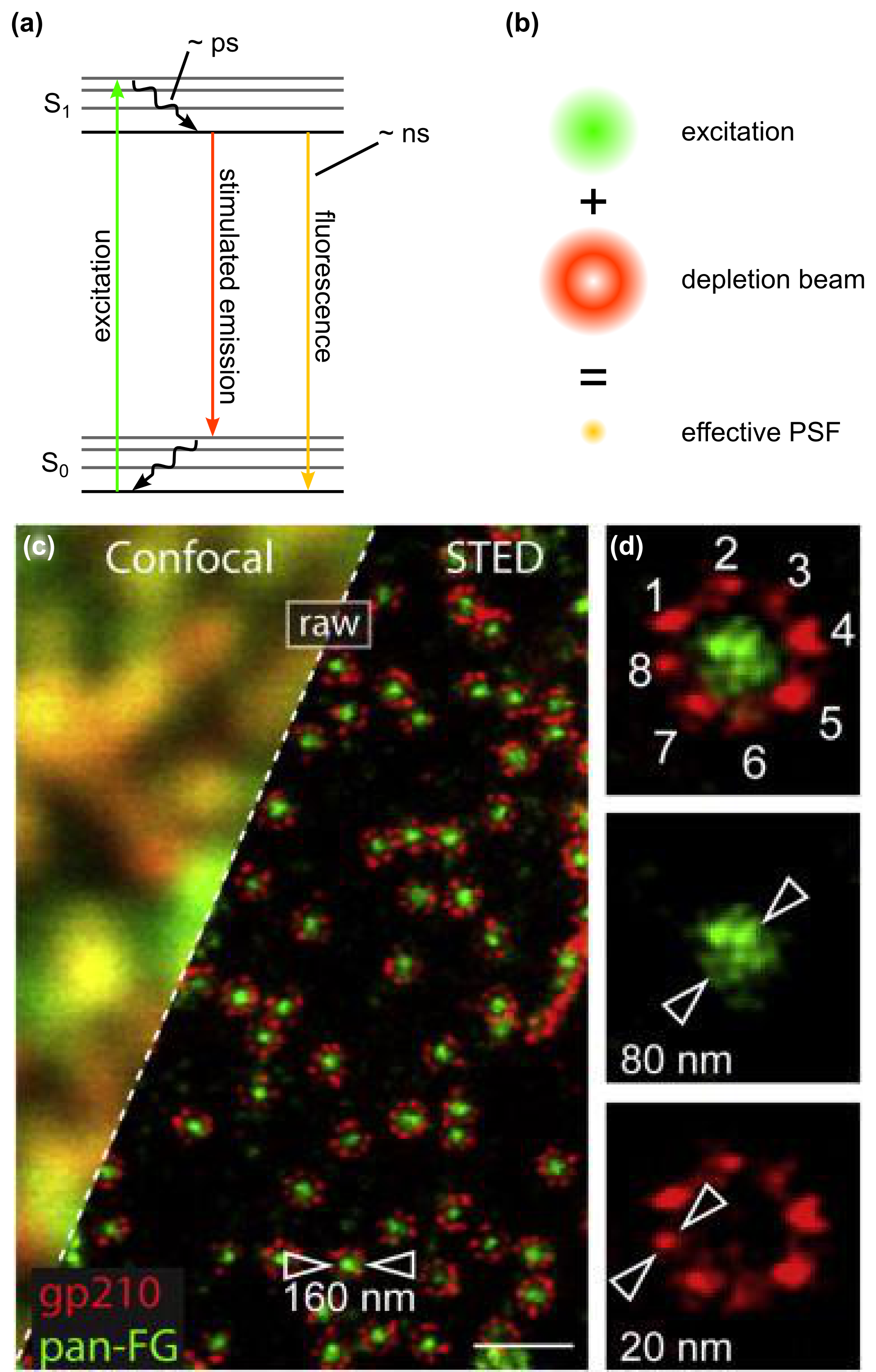}
  \caption{Stimulated emission depletion (STED). {\bfseries (a)} Overview of the photophysics processes involved: Excitation, stimulated emission and fluorescence. {\bfseries (b)} Effective sub-diffraction limited PSF as a result of excitation and doughnut-shaped depletion laser beams. {\bfseries (c)} STED image of immunolabeled subunits in amphibian NPC, raw data smoothed with a Gaussian filter extending over 14 nm in FWHM. Scale bar: 500 nm. {\bfseries (d)} Individual NPC image showing eight antibody-labeled gp210 homodimers. (c,d: Reproduced with permission from \cite{Goettfert2013}.)}
  \label{fig_sted}
\end{figure}

It is important to note that the depletion doughnut beam itself is also diffraction limited, but the effective size of its hole in the middle can be adjusted by the beam intensity. In other words, the higher the intensity of the depletion beam, the farther one goes into the saturation of the fluorophores. As a result of this nonlinear behavior, the fluorescence PSF can be sculpted. It follows that the resulting sub-diffraction resolution can be described by a modified form of Ernst Abbe's equation after considering the degree of saturation,
\begin{equation}
  \label{eqn_hell}
  d = \frac{\lambda}{2 \mathrm{NA} \sqrt{1 + I/I_{\mathrm{sat}}}} \quad .
\end{equation}
Here, $I$ is the peak intensity of the depletion laser and $I_{\mathrm{sat}}$ denotes the saturation intensity of the fluorophore. The resolution becomes sub-diffraction limited when the ratio between $I$ and $I_{\mathrm{sat}}$ becomes larger than one. The best resolution that has been demonstrated with STED is about 20 nm in the case of standard fluorescence labeling assays with organic fluorophores \cite{Donnert2006,Meyer2008,Goettfert2013} and about 50 nm using genetically expressed fluorescent proteins in live cells \cite{Willig2006}. In the case of a very robust fluorophore such as a nitrogen-vacancy color center in diamond, a resolution down to 2-3 nm was successfully demonstrated \cite{Wildanger2012,Arroyo-Camejo2013}.

\subsubsection{The RESOLFT concept}
Sub-diffraction imaging by using a doughnut-shaped excitation pattern for fluorescence suppression can be generalized to other photophysical mechanisms besides stimulated emission. For example, one can exploit the saturated depletion of fluorophores, which can be reversibly switched between a bright state and a dark state. This dark state can have a variety of origins such as the ground state of a fluorophore in STED, its triplet state in ground-state depletion (GSD) microscopy \cite{Hell1995,Bretschneider2007}, or a nonfluorescent isomer of an over-expressed fluorescent protein \cite{Hofmann2005}. This general principle was coined ``RESOLFT'' as an abbreviation for reversible saturable optically linear fluorescence transitions\footnote{To the extent that any saturation process is nonlinear, the emphasis on ``linear'' transitions is somewhat of an unfortunate formulation.}, which is also a pun on the way Germans might pronounce ``resolved'' \cite{Hell2007}.

The RESOLFT concept can also be applied to SIM, also known as saturated SIM (SSIM) \cite{Heintzmann2002}. In this regime, a sinusoidal illumination pattern becomes effectively more and more rectangular as the excitation intensity increases. This leads to higher order Fourier terms in the illumination pattern periodicity $a_i$. Using similar back-of-the-envelope considerations as for standard SIM, $|1 / a_i - 1/a_s| \le k_0$, one can show that spatial frequencies $1 / a_s > 2 k_0$ become detectable. Indeed, using SSIM, lateral spatial resolutions on the order of 50 nm have been demonstrated \cite{Gustafsson2005}.

\subsubsection{Challenges ahead for the RESOLFT methods}
The general class of RESOLFT methods has offered a very clever strategy for circumventing the diffraction limit in fluorescence microscopy. Nevertheless, these methods are accompanied with challenges, which will call for more innovations in the years to come. One of the issues is the fact that so far RESOLFT has been a raster scan technique with a certain temporal resolution, which might limit live cell imaging applications or the study of dynamic processes. This issue can be resolved to some extent by massive parallelization \cite{Bingen2011,Chmyrov2013,Yang2014}, e.g. via a square grid similar to structured illumination microscopy.

A second challenge concerns the photophysics of the fluorophores. Excitation to higher electronic states can efficiently compete with the stimulated emission process, opening pathways for photobleaching \cite{Hotta2010}. The company Abberior has addressed this problem by developing a range of suitable dye molecules and other fluorophores covering the greater part of the visible spectrum \cite{Abberior}.

A further area of development will be multicolor applications as it is common in biological fluorescence microscopy. Since already two lasers are necessary for RESOLFT methods, the implementation of two or more color channels is somewhat more complex than in standard fluorescence microscopy. Nevertheless, STED with two different fluorescence labels and two pairs of excitation and STED lasers has been demonstrated \cite{Donnert2007}. There have also been efforts for sharing the excitation and STED lasers \cite{Tonnesen2011}, or separating a third fluorophore that lies in the same spectral region by its fluorescence lifetime \cite{Bueckers2011}.

Imaging deep in the sample also poses difficulties for RESOLFT microscopy. As the attainable resolution critically depends on the quality of the intensity minimum and phase fronts in the center of the depletion beam, slight changes of the depletion beam profile by scattering deteriorates the imaging performance. This problem can be, in principle, addressed by recent developments using adaptive optics \cite{Gould2012}.

\subsection{Super-resolution microscopy based on single molecule localization}
\subsubsection{Single molecule detection}
In parallel to the developments of SNOM in the 1980s and 1990s, scientists worked hard to detect matter at the level of single ions and single molecules. These efforts were originally more motivated by fundamental issues in spectroscopy and to a good part by the community that had invented methods such as spectral hole burning for obtaining high-resolution spectra in the condensed phase. Although already in the 1970s and 80s there had been indications of reaching single-molecule sensitivity in fluorescence detection \cite{Hirschfeld1976}, the first reliable and robust proof came from the work of W. E. Moerner, who showed in an impressive experiment that a single pentacene molecule embedded in an organic crystal could be detected in absorption at liquid helium temperature \cite{Moerner1989}. Soon after that M. Orrit demonstrated a much better signal-to-noise ratio by recording the fluorescence signal in the same arrangement \cite{Orrit1990}. The ease of this measurement kick-started the field of single molecule fluorescence detection. However, it was not until 1993 that E. Betzig provided the first images of single molecules at room temperature \cite{Betzig1993}. Betzig used a fiber-based aperture SNOM to excite conventional dye molecules on a surface. At this point, there was a strong belief that far-field excitation would not be favorable because it would cause a large background. Interestingly, shortly after that R. Zare's lab demonstrated scanning confocal images of single molecules \cite{Nie1994}. This achievement was the final step towards a widespread use of single molecule fluorescence microscopy.

A particularly interesting feature of fluorescence that was revealed by single molecule detection is photoblinking, i.e. the reversible transition between bright and dark states of a fluorescent molecule. Different physical mechanisms may cause fluorescence intermittencies depending on the type of fluorophore as well as its surroundings \cite{Ambrose1991,Trautman1994,Ambrose1994,Dickson1997}. For example, an excited molecule can undergo a transition to a metastable triplet state with a much longer lifetime than the singlet excited state (see Fig. \ref{fig_triplet}). During this time, the molecule is off because it cannot be excited.
\begin{figure}[htb]
  \centering
  \includegraphics{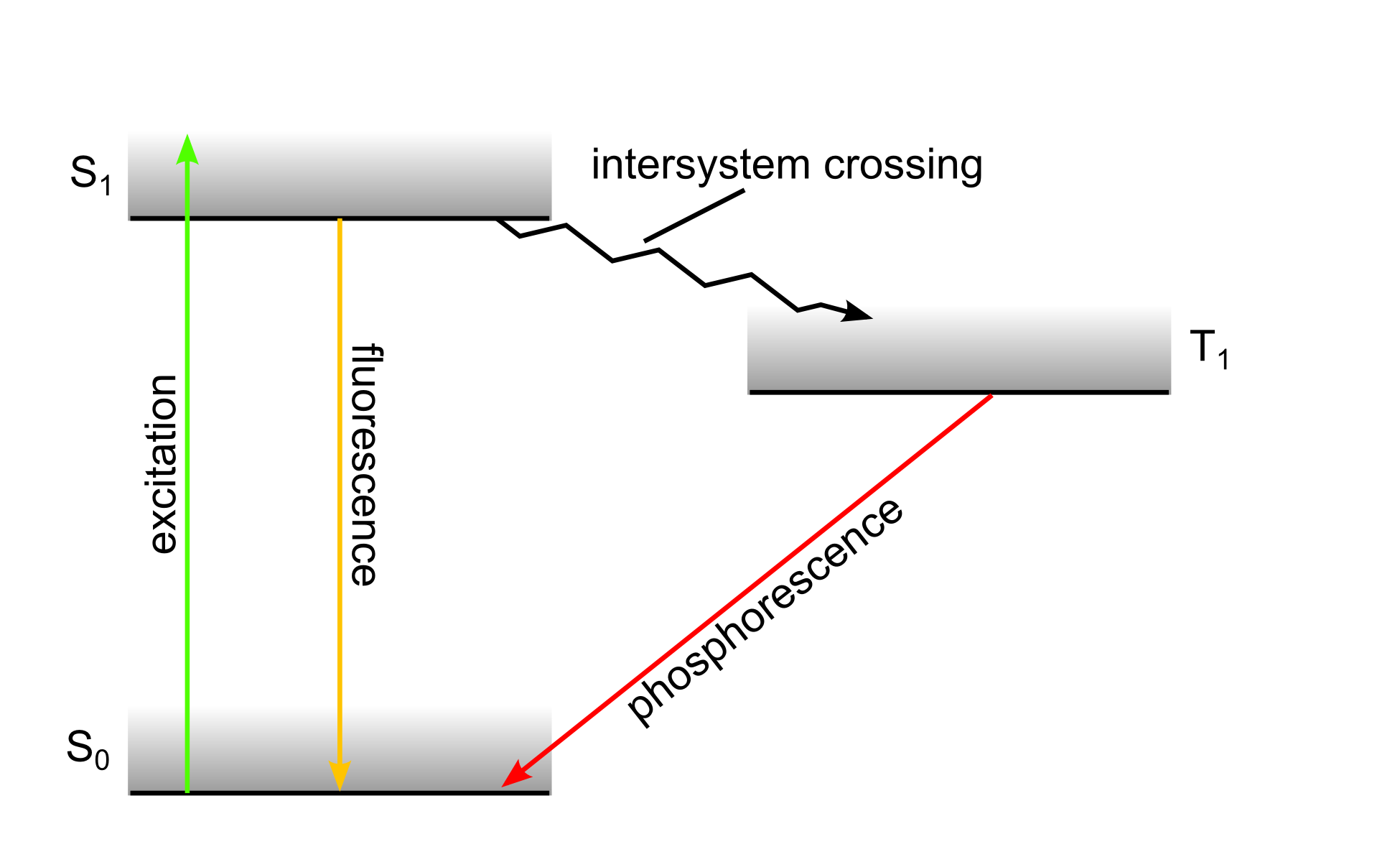}
  \caption{Schematic Jablonski diagram illustrating the reason for fluorescence intermittencies in the emission of a single molecule: Intersystem-crossing leads to the excitation of a dark, long-lived triplet state.}
  \label{fig_triplet}
\end{figure}
The evidence for triplet state blinking is an off-time distribution that follows an exponential law. However, in some cases the off-time statistics reveal a power law similar to the blinking observed in semiconductor nanocrystals \cite{Kuno2001,Shimizu2001}. A proposed mechanism for such a fluorescence intermittency is the formation of a radical dark state \cite{Zondervan2003}. There have been several studies on the topic of blinking, but there is only a limited amount of room-temperature and low-temperature data available and many questions remain open \cite{Suzuki2011,Hoogenboom2007,Schuster2005,Hoogenboom2005,Yeow2006,Orlov2012,Sluss2009}.

Fluorophores also undergo photobleaching, i.e. an irreversible transition to a non-fluorescent product. At room temperature dye molecules typically photobleach within several tens of seconds or a few minutes if sophisticated antifading reagents are used in the buffer. However, the survival times at cryogenic temperatures can go beyond an hour \cite{SigiSPIE2013} or even days in the case of a crystalline matrix. In the special case of terrylene in p-terphenyl a comparable photostability has been achieved even at room temperature \cite{Pfab2004}. Unfortunately, the combination of aromatic molecules and crystalline host matrices is not compatible with the labeling strategies in the life sciences.

\subsubsection{Single molecule localization}
The idea behind localization microscopy is to find the position of each fluorophore by determining the center of its diffraction-limited PSF with a higher precision than its width. This is accomplished by fitting the distribution of the pixel counts on the camera with a model function that describes this distribution (see Fig. \ref{fig_sml}). The principle was already conceived by Werner Heisenberg in the 1920s \cite{Heisenberg1930} and experimentally demonstrated in the 1980s in the context of localizing a single nanoscale object with nanometer precision \cite{Sheetz1988}.

\begin{figure}[htb]
  \centering
  \includegraphics{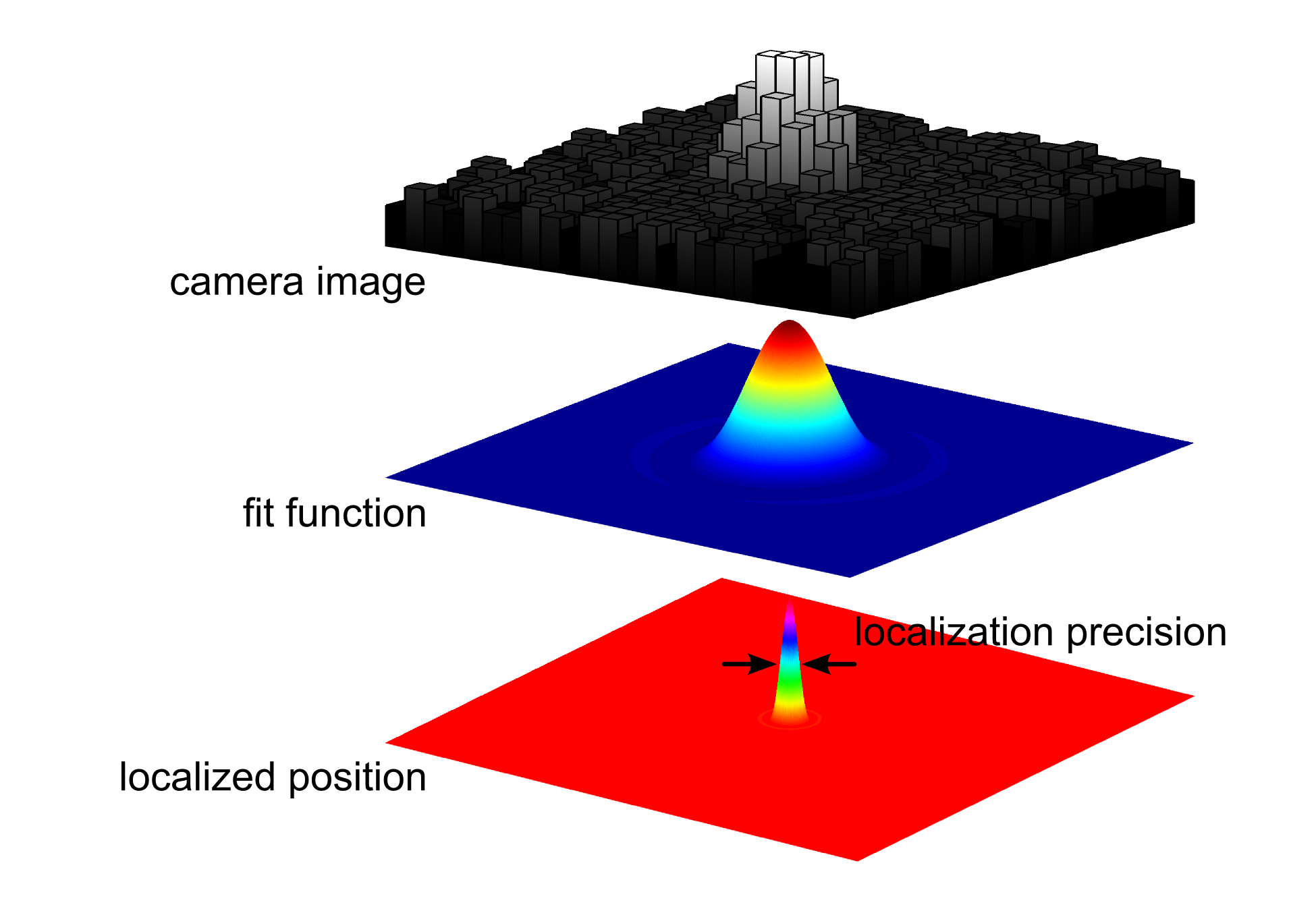}
  \caption{Single molecule localization. The position of single fluorophores can be determined by fitting the distribution of pixel counts on the camera with a model function that describes the distribution. The localization precision depends only on the available SNR.}
  \label{fig_sml}
\end{figure}

In this scheme, single emitters can be localized with arbitrarily high precision only dependent on the available SNR. The localization precision is mainly determined by the number of photons that reach the detector, the size of the PSF, the level of background noise and the pixel size \cite{Thompson2002,Ober2004}. The background noise is in turn affected by the luminescence of the cover slip or other materials on the sample as well as the dark counts and readout noise of the camera \cite{Jung2013}. The attainable localization precision ($\sigma_{\rm{loc}}$) can be written as
\begin{equation} \nonumber\label{equation_mortensen}
\sigma_{\rm{loc}}  = \sqrt{ \frac{s^2 + a^2/12}{N} \left( \frac{16}{9} + \frac{8\pi(s^2 + a^2/12)b^2}{Na^2}\right)} \quad,
\end{equation}
using a maximum likelihood estimation procedure with a 2D Gaussian function. This predicts a localization error close to the information limit \cite{Mortensen2010}. Here, $N$ denotes the detected number of photons, $s$ stands for the half-width of the PSF given by the standard deviation of a Gaussian profile, $b$ is the level of background noise and $a$ denotes the pixel size. The limiting factor is typically the finite value of $N$ caused by irreversible photobleaching of the fluorophore. The photon budget of commonly used photoactivatable fluorescent proteins lies in the range of a few hundred detected photons \cite{McKinney2009}, which typically leads to a localization precision on the order of 20 nm. To improve on this limitation, several efforts have optimized the choices of fluorophores and the buffer conditions \cite{Yildiz2003,Pertsinidis2010}, engineered the dye molecule itself \cite{Vaughan2012}, or carefully controlled its environment \cite{Lee2013}. The best localization precision for single molecules that has been reported is just under 0.3 nanometers \cite{SigiSPIE2013}.

A particularly powerful tool based on the concept of localization is single particle tracking. Localizing a fluorescent marker or non-fluorescent nano-object of interest as a function of time allows one to study dynamical processes such as diffusion in lipid membranes \cite{Saxton1997}. Single particle tracking has been performed with various imaging modalities including fluorescence \cite{Sheetz1983,Schuetz1997}, scattering \cite{Fujiwara2002,Kusumi2005} and absorption \cite{Lasne2006}. However, high temporal and spatial precisions call for a trade-off because smaller integration times lead to a lower signal and lower SNR. Interferometric scattering microscopy (iSCAT) can provide an ideal solution \cite{Hsieh2014}, offering up to MHz frame rates in combination with nanometer localization precision in the case of small scatterers like gold nanoparticles.

Interestingly, localization techniques and particle tracking have also found applications in a wide range of studies such as coherent quantum control \cite{Gorshkov2008} or cold atoms \cite{Bakr2009,Sherson2010}. For example, identification of atoms down to the single lattice site of an optical lattice has provided an avenue to manipulating single qubits and studying many-body effects like the quantum phase transition from a superfluid to a Mott insulator.

\subsubsection{Co-localization microscopy}
Given that a single molecule can be localized to an arbitrarily high precision, one could also resolve two nearby molecules if only one could address them individually. This was formulated by E. Betzig in 1995 as a general concept \cite{Betzig1995}, but it was already demonstrated experimentally in 1994 by the group of Urs Wild in cryogenic studies \cite{Guettler1994}. In the latter, the inhomogeneous distribution of the molecular resonance lines allows one to address each molecule separately by tuning the frequency of a narrow-band excitation laser. We have recently demonstrated that the same principle of spectral selection can also be used to address single ions in a solid \cite{Utikal2014}. By combining cryogenic high-resolution spectroscopy and local electric field gradients, it was also shown in our laboratory that two individual molecules could be three-dimensionally resolved with nanometer resolution \cite{Hettich2002}. To our knowledge, those results still establish the highest three-dimensional optical resolution.

Several groups have tried different strategies for distinguishing neighboring fluorophores. One example was to use semiconductor nanocrystals with different emission spectra \cite{Lacoste2000} or stepwise bleaching of single molecules \cite{Gordon2004,Qu2004}. However, extension of these methods to very large number of fluorophores was not practical. The decisive breakthrough came in 2006, when three groups reported very similar strategies based on stochastic photoactivation processes that switched the fluorophores between a dark state and a fluorescent state (see also Fig. \ref{fig_storm}b). Eric Betzig and colleagues called their method photoactivated localization microscopy (PALM) \cite{Betzig2006}, Sam Hess and his team coined the term fluorescence photoactivation localization microscopy (FPALM) \cite{Hess2006}, and X. Zhuang and her group used the term stochastic image reconstruction microscopy (STORM) \cite{Rust2006}. In each case, the fluorophores are placed on the target structure by different labeling techniques. While commonly used antibody-based assays have a label-to-target distance of about 20 nm, using nanobodies \cite{Ries2012}, aptamers \cite{Opazo2012} or fluorescent proteins \cite{Terry1995} can reduce that distance to a few nanometers.

\begin{figure}[p]
  \centering
  \includegraphics{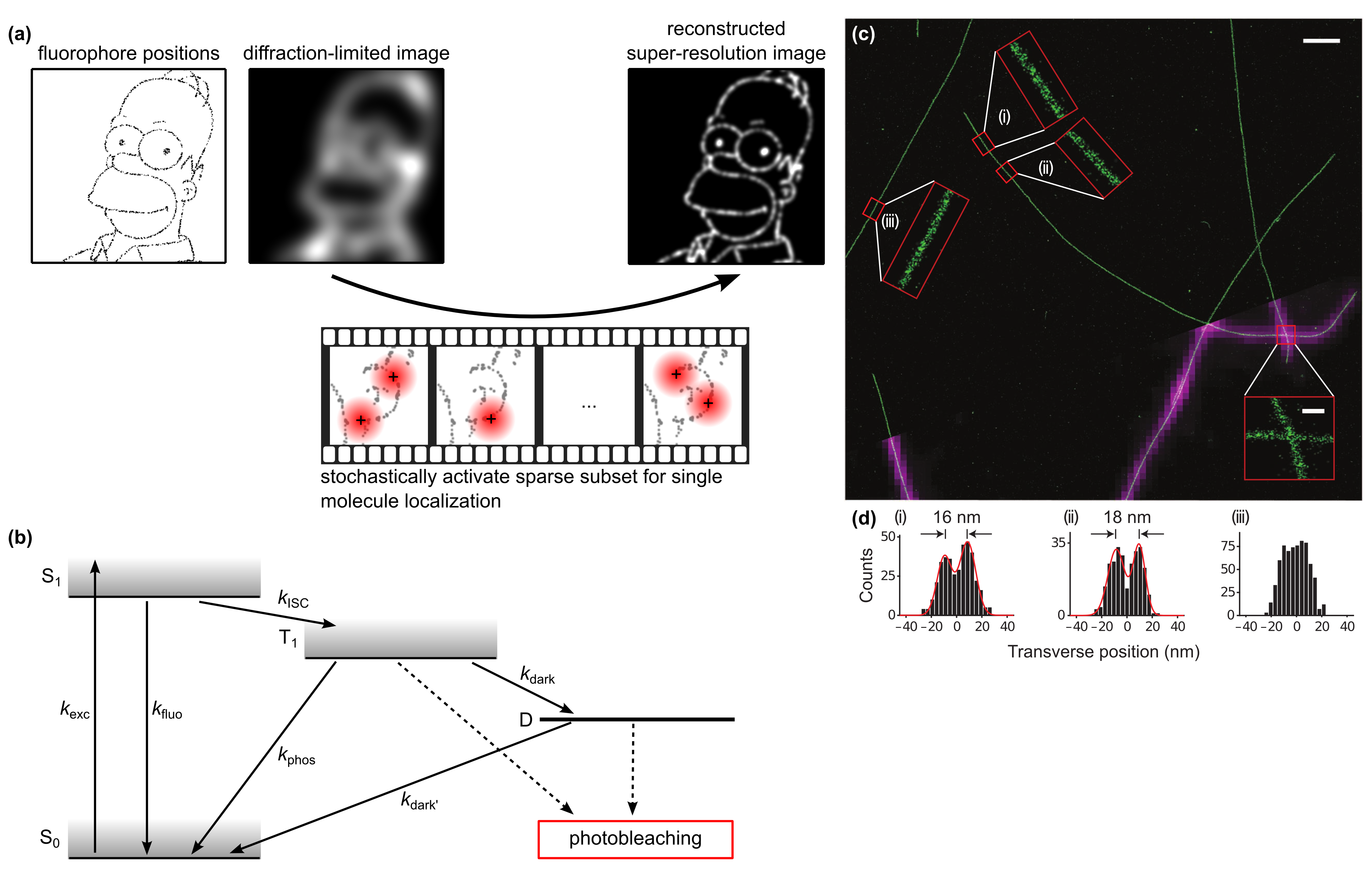}
  \caption{Super-resolution imaging based on single molecule localization. {\bfseries (a)} Illustration of the image acquisition and reconstruction procedure. The target structure is described by the positions of the fluorescence labels. Details in the structure cannot be recognized by diffraction-limited imaging. Repeated stochastic activation, imaging and localization of a sparse subset of fluorophores allows one to reconstruct a super-resolution image. {\bfseries (b)} Schematic energy diagram for a generic photo-switchable fluorescent molecule as used for example in STORM. {\bfseries (c)} A STORM image of microtubules (green) with several magnified images shown in the insets. A portion of the corresponding conventional fluorescence image (magenta) is overlaid on the STORM image. {\bfseries (d)} Segments showing hollow microtubule profiles with inner diameters of 16 - 18 nm. The red curves are nonlinear least-square fits of the distribution to two Gaussian functions. Scale bars: 1 $\mu$m, main image; 100 nm, insets. (c,d: Reproduced with permission from \cite{Vaughan2012}.)}
  \label{fig_storm}
\end{figure}

Figure \ref{fig_storm} illustrates the data acquisition procedure for super-resolution imaging based on single molecule localization. By shining light on the sample with a blue-shifted activation laser beam, one can stochastically switch on a sparse subset of fluorophores. Next, one turns on the excitation laser and collects the fluorescence from the few activated fluorescent molecules until they become deactivated. By adjusting the intensity of the activation beam, one can control the average number of activated fluorescent labels to ensure that PSFs from individual fluorophores do not overlap. One then performs a localization analysis for each recorded PSF to determine the positions of all molecules. The process of activation, recording and localization is then repeated for many other random subsets of fluorophores until one is satisfied with the number of labels for reconstructing a super-resolution image. There are also variations of this acquisition procedure using, for instance, asynchronous activation and deactivation of fluorophores \cite{Sharonov2006} or assays where diffusing fluorophores get activated upon binding to the target structure \cite{Jungmann2010,Schoen2011}.

In the standard super-resolution imaging modalities such as PALM and STORM, usually photochemistry is employed in order to exert some degree of control on the photoswitching kinetics. This control is necessary since the achievable resolution also depends on the ratio of the on- and off-switching rates of the used fluorophores \cite{vandeLinde2010}. Examples of such photochemistry is the chromophore cis-trans isomerization or protonation change in the case of fluorescent proteins \cite{Bourgeois2012}, and the interplay of reduction and oxidation using enzymatic oxygen-scavenging systems and photochromic blinking for organic dye molecules \cite{Vogelsang2008,Ha2012}.

The state of the art in resolution for localization microscopy is about 10 nm \cite{Vaughan2012} limited by the total number of photons emitted before photobleaching. However, even sub-nanometer localization precision has been reported in cases where photobleaching could be delayed by using oxygen scavengers \cite{Pertsinidis2010} or cryogenic conditions\cite{SigiCPC2014}. The latter measurements offer the additional advantage of a more rigid sample fixation than chemical fixatives. Indeed, first studies exploring PALM imaging at low temperature have also recently surfaced \cite{Chang2014,Kaufmann2014}. The full arsenal of methodologies developed for cryogenic electron microscopy may be applied to prepare samples for cryogenic super-resolution microscopy and even dynamic processes could be studied either by stopping processes at different times or by employing methods like local heating with an infrared laser \cite{Zondervan2006}.

\subsubsection{Challenges ahead in co-localization microscopy}
A quantitative assessment of the molecules' positions critically depends on both the precision and accuracy of the employed method. Precision in localization microscopy is determined by the standard deviation of the estimated position of an emitter assuming repeated measurements, whereas the accuracy quantifies how close the estimated position lies to the true position. In other words, even if the measurement precision is high, absolute distance information might be compromised by technical sources of bias like pixel response non-uniformity of the camera or sample drift \cite{Pertsinidis2010}.

An important systematic source of error concerns the dipolar emission characteristics of single molecules. It is known that the image of an arbitrarily oriented dipolar light source deviates from a simple isotropic PSF (see Fig. \ref{fig_dipoleorientation}). As a result, localization of the individual fluorophores requires a fitting procedure that takes this effect into account \cite{Bartko1999,Enderlein2006}. In the presence of nearby interfaces, this can become a nontrivial task \cite{SigiCPC2014}. However, the PSF asymmetry is much less pronounced if a microscope objective with low numerical aperture is used \cite{Enderlein2006}. Of course, fitting the data with full theoretical treatment of the PSF or good approximations provides accurate values for the position and orientation of the fluorophore even in the case of high numerical aperture \cite{Aguet2009,Mortensen2010,Stallinga2012}.

The most severe limit in localization microscopy is the difficulty of high-density labeling. Here it is to be remembered that the image in this method is constructed by joining the centers of the individual fluorophores (see Fig. \ref{fig_storm}). This means that a resolution of 5 nm in deciphering the details of a figure would need at least two fluorophores that are spaced by about 2.5 nm according to the sampling theorem \cite{Shannon1949}. The first problem with this requirement is the difficulty of labeling at such high densities. Second, once one manages to place the fluorophores at the right place, one faces the problem that such closely-spaced fluorophores undergo resonance energy transfer (homo-FRET) \cite{Foerster1948,Luchowski2008}. As a result, the emission cannot be attributed to one or the other fluorophore, and the basic concept of localization microscopy breaks down.

Another issue to be considered is that of accidental overlapping PSFs from neighboring fluorophores. In high-precision co-localization microscopy this situation can usually be avoided by rejecting the affected PSFs. However, it would be more advantageous to localize these PSFs as well. It turns out that the fluorophore density can be increased to achieve faster acquisition times by using appropriate algorithms \cite{Holden2011,Huang2011}.

\begin{figure}[htb]
  \centering
  \includegraphics{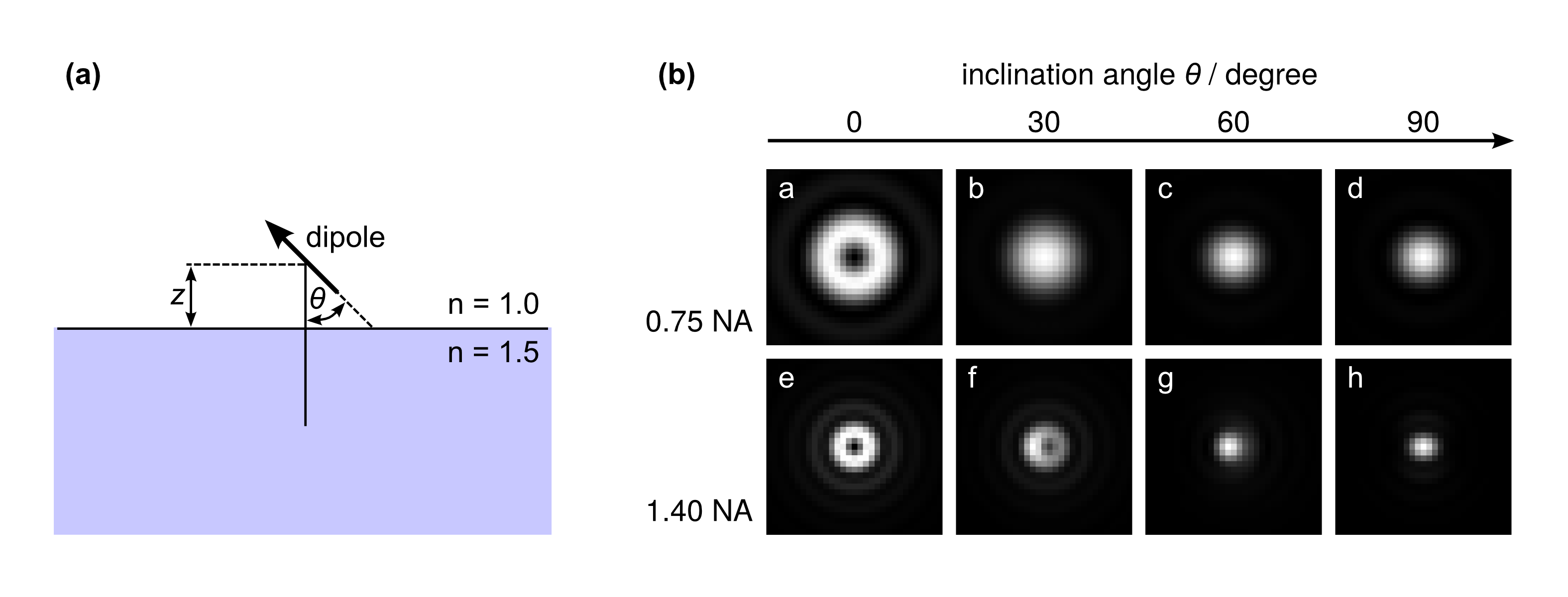}
  \caption{Simulations of PSFs of dipoles near an interface. {\bfseries (a)} Geometry for the simulations. The molecules have an inclination angle $\theta$ and the distance between the molecule and the interface is $z$. {\bfseries (b)} Examples for PSFs for two different numerical apertures and four inclination angles at $z$ = 2 nm.}
  \label{fig_dipoleorientation}
\end{figure}

\subsection{State-of-the-art resolution and ongoing efforts in super-resolution microscopy}

The lateral resolution in the above-mentioned super-resolution methods is currently of the order of 10 - 50 nm, but these techniques are not fundamentally limited by any particular physical phenomenon. The resolution is rather hampered by practical issues, which can be addressed to various degrees in different applications. Thus, only emphasizing a record resolution outside a specific context is not a meaningful exercise.

An important development concerns super-resolution in three dimensions. One possibility to localize a fluorophore along the optical axis is via astigmatism \cite{Huang2008}. By inserting a cylindrical lens in the detection path, the PSF becomes elliptical, from the degree of its ellipticity and orientation one can deduce the additional axial position of the fluorophore. Lateral localization precision of about 25 nm with an axial localization precision of about 50 nm was already reported in 2008 \cite{Huang2008,Huang2008b}. Recently, an isotropic localization precision of about 15 nm in all three spatial dimensions was reported using STORM in combination with an Airy-beam PSF \cite{Jia2014}. An alternative way to obtain 3D super-resolution is multi-focal plane imaging \cite{Juette2008,Abrahamsson2013}. Here, different focal planes are imaged on various regions of the camera by splitting the fluorescence light and introducing different path lengths. The height can then be deduced from the degree of defocusing. Another approach uses an engineered PSF that encodes the axial position of the emitter in the rotation angle of two lobes, a double-helix PSF \cite{Pavani2009}.

A crucial requirement for practical biological microscopy is the ability to image many entities simultaneously. The most convenient and common approach is to label different parts of the specimen with fluorophores of distinct absorption or emission spectra. Considering that the new super-resolution methods, including RESOLFT, PALM, STORM, etc., all rely on the photophysical properties of the fluorophores, it is not a trivial matter to marry them with multicolor imaging. First, fluorescence probes with the desired switching properties must be available with the correct excitation, emission and activation wavelengths. Interestingly, scientists began to develop multicolor super-resolution solutions shortly after the introduction of localization microscopy. Indeed, two-color \cite{Bates2007,vandeLinde2009,Lampe2012} and even three-color super-resolution imaging has been demonstrated \cite{Bossi2008}. A second challenge concerns the crosstalk that is caused by the spectral overlap of the emission bands of different fluorophores, which are typically several tens of nanometers broad at room temperature. A possible solution would be to perform super-resolution microscopy at low temperatures because even though dye molecules suited for labeling in life science do not show lifetime-limited linewidths at cryogenic temperatures, their spectra can become narrower by orders of magnitude.

As super-resolution optical microscopy becomes a workhorse, it becomes more and more important that the methods can also handle live cell imaging. Here, imaging speed and phototoxicity pose important problems. On the one hand, fast imaging often requires a large excitation dose to be able to collect lots of photons in a short time window. High light dose, however, causes the production of free radicals through the photo-induced reaction of the fluorophore with molecular oxygen \cite{Dixit2003}. Furthermore, excess light also brings about fast photobleaching and short observation times. An interesting approach to minimize these phototoxic effects is light-sheet microscopy \cite{Huisken2004}. Here, the wide-field detection is disentangled from the illumination, which consist of a thin sheet of light perpendicular to the focal plane. By performing tomographic recordings at different sample orientations, one can then obtain impressive three-dimensional images of whole organs in small animals such as zebrafish \cite{Mickoleit2014}. Of course, diffraction limits the thickness of the light sheet to dimensions well above a wavelength, especially if the illumination area is to be large. By employing slowly diffracting beams such as Bessel beams, one can minimize the problem of beam divergence \cite{Planchon2011}. Light-sheet microscopy is very popular in developmental biology, where super-resolution is less important than large-scale information about the whole system over a longer time. An application example of the technique is the four-dimensional imaging of embryos at single-cell resolution \cite{Krzic2012,Tomer2012}.

If we now relax the requirements for routine biological microscopy, we find that several experiments have already extended super-resolution microscopy to the nanometer and even sub-nanometer level. The first demonstration of nanometer resolution in all three spatial dimensions used low-temperature fluorescence excitation spectroscopy in combination with recording the position-dependent Stark shift of the molecular transition in an electric field gradient \cite{Hettich2002}. In another experiment, a distance of about 7 nm was measured between two different dye molecules with an accuracy of 0.8 nm using a feedback loop for the registration of two color channels and oxygen-reducing agents\cite{Pertsinidis2010}.

\begin{figure}[htb]
  \centering
  \includegraphics{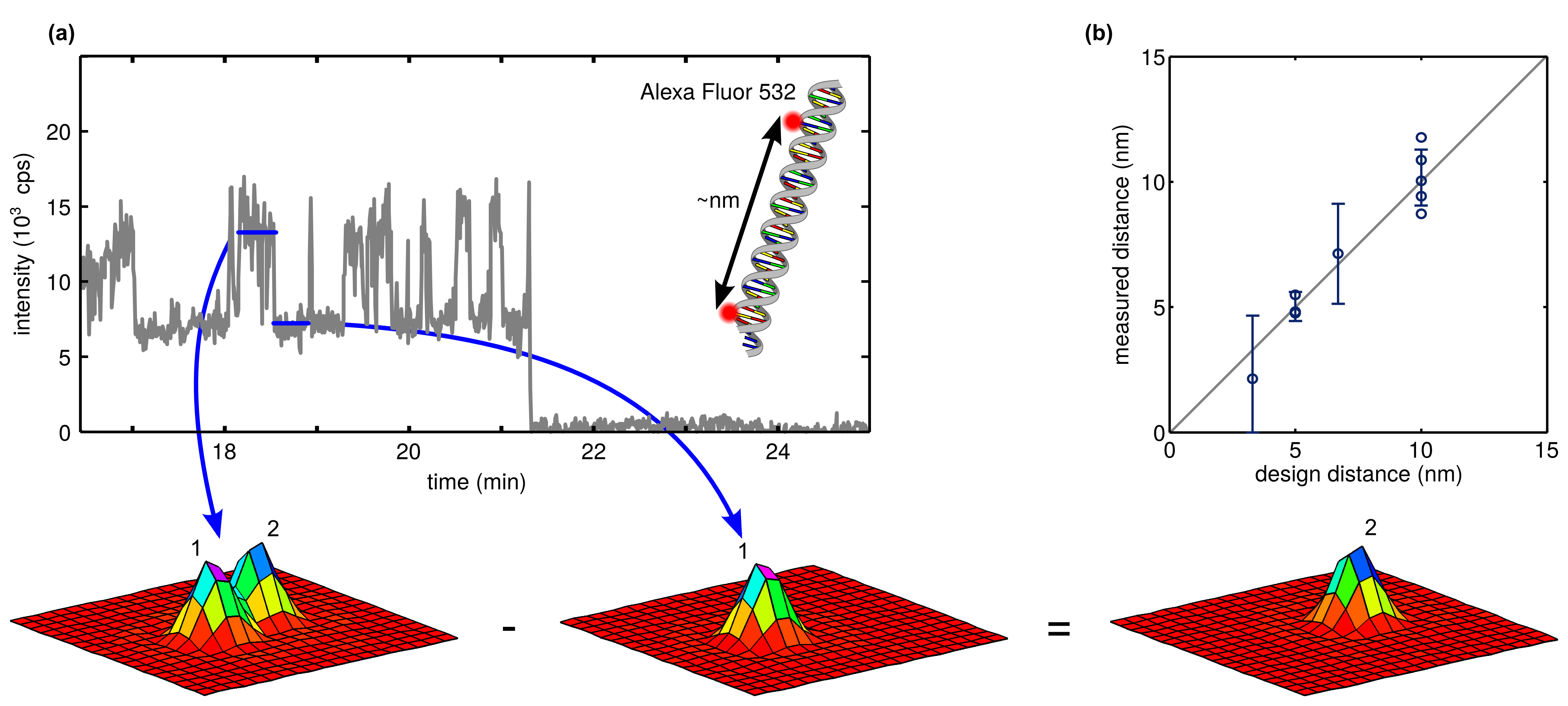}
  \caption{Single molecule colocalization using photoblinking. {\bfseries (a)} Fluorescence intensity trace from a DNA molecule labeled with two dye molecules. With the additional information of how many fluorophores are in the bright state at a given time, the positions of both fluorescent molecules and their distance can be computed. Inset: Schematic of the used sample. Two identical fluorophores attached to a double-stranded DNA at well-defined separations. {\bfseries (b)} Using cryogenic colocalization microscopy, the distance between the two fluorescence labels attached to a DNA can be determined with sub-nanometer accuracy.}
  \label{fig_coloc}
\end{figure}

The most recent achievement concerns Angstrom accuracy in cryogenic colocalization. Here, two identical fluorophores attached to a double-stranded DNA at well-defined separations as small as 3 nm were resolved with a distance accuracy better than 1 nm \cite{SigiCPC2014}. Figure \ref{fig_coloc}a illustrates how this can be used to determine the positions of the dye molecules and their separation. The intensity trace retrieved from the image stack of a DNA molecule with the attached fluorescent molecules shows three levels. The upper level corresponds to the state where both fluorophores are on, the middle one to the case that only one is fluorescent, and in the lowest level both fluorophores are in a dark state. After localizing the single fluorophores in the frames where there is only one present, one can subtract the image frames from those where both fluorophores were on. In this fashion, one can also localize the second fluorophore. An alternative route is to find the center of mass of each fluorophore and then compute the distance. Using the latter approach, both sub-nanometer localization precision and accuracy for the distance measurement have been demonstrated (see Fig. \ref{fig_coloc}b).

The holy grail of super-resolution microscopy is to break free from the shackles of fluorescent labels. The use of fluorescence markers introduces a variety of difficulties, starting with the labeling process itself and ending with the inevitable photobleaching of the labels. In the past years, there have been several proof-of-principle demonstrations to obtain fluorescence-free super-resolution images. One possibility is a technique called optical diffraction tomography (ODT) \cite{Sentenac2006,Belkebir2006}. Here, the sample is illuminated using every possible angle of incidence allowed by the numerical aperture of the microscope objective. Then the intensity, phase and polarization state of the scattered far field are recorded for different angles, and the distribution of the permittivity of the object of interest is reconstructed numerically. Recently, an optical resolution of about one-fourth of the wavelength was experimentally demonstrated \cite{Zhang2013}. Label-free super-resolution has also been demonstrated using surface-enhanced Raman scattering (SERS) \cite{Ayas2013}, where one performs a stochastic reconstruction analysis on the temporal intensity fluctuations of the SERS signal. Another intriguing method employs ground-state depletion of the charge carriers with a doughnut shaped beam resulting in the transient saturation of the electronic transition by using a pump-probe scheme \cite{Wang2013}. There have also been several approaches to achieve super-resolution imaging using the photon statistics of the emitters \cite{Schwartz2013,Cui2013}.

As discussed earlier, it is also possible to detect unlabeled biomolecules such as proteins via iSCAT detection of their Rayleigh scattering \cite{Piliarik2014}. In this method the image of a single protein can be localized in the same manner as in Fig. \ref{fig_sml}. Here too, one needs to turn the proteins on and off individually if one wants to resolve them beyond the diffraction limit. In dynamic experiments, the arrival time of each protein can serve as a time tag \cite{Piliarik2014}. The localization precision and therefore the attainable resolution is determined by the signal-to-noise ratio, which in turn depends on the size of the biomolecule in iSCAT. This size-dependent signal also provides a certain level of specificity that in fluorescence modalities can only be achieved by employing different fluorophores.

Another report has used the saturation of scattering in plasmonic nanoparticles. Although the saturation effect in the absorption of small plasmonic nanoparticles has been studied for many years \cite{Link2000}, saturation of scattering has only been reported recently \cite{LeeH2013,Chu2014b}. This saturation stems from a depletion of the plasmon resonance. Similar to the case of super-resolution imaging in SSIM, the saturation effect in scattering allows one to record images with a resolution beyond the diffraction-limit. By recording images at different light intensities, super-resolved images were obtained and a resolution of $\lambda / 8$ was demonstrated \cite{Chu2014}.

Finally, combination of optical microscopy with other imaging modes such as scanning probe techniques or electron microscopy can offer very useful additional information about the sample. Some of the recent examples of correlative microscopy are the combination of optical super-resolution microscopy with electron microscopy \cite{Watanabe2011,Kopek2012,Chang2014} and with atomic force microscopy (AFM) \cite{Harke2012,Chacko2013,Monserrate2013}.

\section{Concluding remarks}
The quest for inventing new imaging mechanisms and pushing the spatial and temporal resolutions is a fundamental challenge for physicists and of great practical importance in science and technology. Abbe's formulation of the diffraction limit at the end of the nineteenth century put a harsh spell on optical microscopy, which lasted for about one hundred years. The advent of scanning near-field microscopy broke this spell, and once the dogma of a fundamental limitation of resolution was eliminated, scientists reconsidered many scenarios and explored fascinating techniques, which we have discussed in this review article.

It is now fully accepted that resolving two small objects at very close distances could be, in principle, achieved with an arbitrary resolution and accuracy. The key concept in breaking the diffraction barrier has been to exploit more information from the system, e.g. taking advantage of the spectroscopic energy levels or transitions of fluorophores. In this spirit, scientists continue to develop optical imaging methods by devising clever schemes that rely on nonlinear phenomena, quantum optics or ultrafast laser spectroscopy.

The many elegant ideas do not, however, all have practical implications. In particular, one has to consider many restrictions in biological imaging. For example, the amount of laser power that one can shine onto a live cell before it is damaged is orders of magnitude lower than what a diamond sample can take. Moreover, there are important issues concerning labeling techniques and the influence of the label on the functionality of its environment. One subtle point regards the production of free radicals in a photochemical reaction of the excited fluorophore with the surrounding oxygen molecules. To minimize the effect of phototoxicity, it is helpful to acquire images as efficiently as possible. Of course, this is also highly desirable because one gets access to more of the dynamics of the biological and biochemical processes.

The Nobel Prize in Chemistry in 2014 has honored the contributions of optical scientists in the area of super-resolution fluorescence microscopy and single-molecule detection. This is the fourth prize after dark-field microscopy (Zsigmondy 1925), phase contrast microscopy (Zernike 1953), and the green fluorescent protein (Chalfie, Shimomura and Tsien 2008), which is dedicated to optical microscopy, spread over nearly one hundred years. The recent achievements are a testimony to the livelihood of light microscopy as a research field in fundamental science. They show a new trend against the older belief that the physics of imaging is fully understood and that its development belongs to engineering departments.

We are convinced that the combination of concepts from laser spectroscopy, quantum optics, photophysics, photochemistry, nanotechnology, and biophysics will introduce many new avenues for optical imaging. Currently resolution at around 10-50 nm is routinely reported in various configurations, but we have shown that this limit can be pushed by another one hundred times to the sub-nanometer level. More importantly various imaging contrasts, e.g. label-free techniques, promise to open the door to whole new classes of information. Finally, temporal resolution will be an area of innovation and growth. The dynamics of interest in biomedical processes range from femtosecond for electronic and vibrational degrees of freedom to days and years for growth and disease progression. This astronomical span of the time scales will certainly require totally different techniques.

Progress in all these cases will ultimately confront the {\em signal-to-noise barrier}. Development of methods for efficient collection of photons and their combinations with lab-on-chip solutions, advances in photochemistry and photophysics of new labels as well as better detector and laser technologies will all contribute to pushing this barrier. Experimental physicists and in particular optical scientists are well positioned to lead the ongoing revolution of optical microscopy if they manage to achieve a high degree of cross fertilization between biology, chemistry, medicine and physics.

\bigskip

\section*{Acknowledgements}
  \ifthenelse{\boolean{publ}}{\small}{}
We thank Anna Lippert, Richard W. Taylor, Simon Vassant and Luxi Wei for discussions and critical comments on the manuscript. We acknowledge support from the Alexander von Humboldt foundation and the Max Planck Society.

\newpage
{\ifthenelse{\boolean{publ}}{\footnotesize}{\small}
 \bibliographystyle{osajnl}  
  \bibliography{sigi-vahid-textcp} }     

\ifthenelse{\boolean{publ}}{\end{multicols}}{}

\end{bmcformat}
\end{document}